%% file: main.tex
\documentclass[smallextended]{article}
\usepackage{graphicx,amsmath,amssymb,hyperref,cleveref,xcolor,subcaption,makecell,booktabs,comment,svg,enumitem,mathtools,tikz-cd} 
\usepackage[margin=1in]{geometry}
\usepackage[font={small},labelfont={bf}]{caption}
\usepackage{tcolorbox}
\tcbuselibrary{breakable}
\usepackage[round,numbers,sort&compress]{natbib}
\usepackage[margin=1in]{geometry}
\input{macros}

\usepackage[normalem]{ulem}
\usepackage[rgb,dvipsnames]{xcolor}
\usepackage{bm}
\usepackage{booktabs}

\newif\ifshowchanges
 \showchangesfalse    % changes appear black

\ifshowchanges
    \newcommand{\changed}[1]{\textcolor{blue}{#1}}
\else
    \newcommand{\changed}[1]{#1}
\fi

\begin{document}

\begin{center}
\rule{\textwidth}{0.4pt}\\[1em]
    {\Large \textbf{A multiscale discrete-to-continuum framework for structured population models}}
\rule{\textwidth}{0.4pt}\\[1em]
    {\large Eleonora Agostinelli$^1$* \hspace{0.1cm} \textbullet \hspace{0.1cm} Keith L. Chambers$^{1,2}$ \hspace{0.1cm} \textbullet  \hspace{0.1cm} Helen M. Byrne$^{1,2}$ \hspace{0.1cm} \textbullet \hspace{0.1cm} Mohit P. Dalwadi$^{1}$ } \\[1em]
    {$^{1}$Mathematical Institute, University of Oxford, OX2 6GG Oxford, UK}\\[1em]
    {$^{2}$Ludwig Institute for Cancer Research, University of Oxford, OX3 7DQ Oxford, UK }\\[1em]
    {\small \texttt{*eleonora.agostinelli@maths.ox.ac.uk}} \\[2em]
\end{center}

\begin{abstract}
Mathematical models of biological populations commonly use discrete structure classes to capture trait variation among individuals (e.g. age, size, phenotype, intracellular state). Upscaling these discrete models into continuum descriptions can improve analytical tractability and scalability of numerical solutions. Common upscaling approaches based solely on Taylor expansions may, however, introduce ambiguities in truncation order, uniform validity and boundary conditions. To address this, here we introduce a discrete multiscale framework to systematically derive continuum approximations of structured population models. Using the method of multiple scales and matched asymptotic expansions applied to discrete systems, we identify regions of structure space for which a continuum representation is appropriate and derive the corresponding partial differential equations. The leading-order dynamics are given by an advection equation in the bulk domain and advection-diffusion processes in small inner layers about wavefronts. We further derive discrete boundary layer descriptions for regions where the discrete model cannot be approximated by a continuum description. Finally, we demonstrate the method on a simple lipid-structured model for early atherosclerosis and verify consistency between the discrete and continuum descriptions. The multiscale framework we present can be applied to other heterogeneous systems with discrete structure in order to obtain appropriate upscaled dynamics with asymptotically consistent boundary conditions. 
\end{abstract}

\section{Introduction}

Biological organisation is commonly underpinned by structure, which refers to organised heterogeneity, from the population-level (e.g. age, size) to phenotypic differences within cellular environments. This heterogeneity shapes individual interactions, development, and population dynamics. Structured mathematical models provide a framework for representing biological heterogeneity by introducing internal state variables into the population description. 
\changed{Depending on the nature of the heterogeneity and the modelling assumptions, these variables may be represented continuously \cite{lorenzi2025phenotype} or partitioned into discrete states or groups \cite{robertson2018matter,walker2025discretely}. The use of a discrete or continuous representation is a modelling choice guided by the biology of the system and the question being asked, rather than one type of representation being correct in general. For example, age has been represented using both discrete \cite{ram2021modified} and continuous \cite{abo2026survival} structure variables. Similarly, both discrete and continuous representations have been used to describe phenotypic \cite{crossley2024phenotypic,celora2023spatio} and size-based heterogeneity \cite{chambers2023new,ford2019lipid}.}
Discrete structured models are used in various fields, including epidemiology (age-based disease transmission analysis) \cite{tudor1985age,griffiths2000age,ram2021modified}, cell biology (cell cycle transitions, phenotype switching, exhaustion, migration) \cite{vittadello2019mathematical,eftimie2021mathematical,lai2025impact,crossley2025modelling}, atherosclerosis (lipid metabolism and accumulation, and macrophage behaviour) \cite{cobbold2002lipoprotein,ford2019efferocytosis,chambers2023new,chambers2024blood,chambers2025spatially}, and agent-based models, e.g. \cite{chaplain2020bridging,stace2020discrete,kory2024discrete,wattis2020mathematical,simpson2024discrete,crossley2025modelling,pillay2017modeling}.

Mathematically, structured heterogeneity in discrete frameworks is often represented by a large system of coupled equations, which can be challenging to analyse directly. Continuum approximations typically recast the dynamics in terms of partial differential equations (PDEs), which can be more amenable to investigation by a wider range of analytical techniques. Additionally, \changed{while the size of a discrete system scales with the number of structure classes, the continuous PDE system need only be resolved according to the smoothness of the structure distribution itself, allowing more scalable numerical solutions.} Discrete-to-continuum approaches which retain the essential features of the original system are a central tool in structured models \cite{barry2020discrete}, but as we shall see there is considerable nuance in how continuum limits should be taken \cite{walker2025discretely}. \changed{While the discretisation of continuum models is well established, the reverse process of systematically deriving continuum descriptions from discrete structured models remains less explored.}

\changed{In this paper, we take a discrete model as our starting point, reflecting the many settings in which this is the natural modelling choice. We ask when a consistent continuum approximation can be derived from this discrete model, and how to derive this systematically. We do not address when a discrete or continuum model is more suitable to describe the underlying biology in general; our aim is to connect the two mathematical descriptions systematically. }

A standard method for obtaining continuum approximations involves using a Taylor series expansion (see e.g. \cite{chambers2023new,chaplain2020bridging,simpson2024discrete,fozard2010continuum,murray2012classifying,barry2022continuum,kiradjiev2023multiscale,kory2024discrete2}). The discrete population is represented by a continuous density in the limit of many structure classes and a Taylor series expansion is applied to the discrete system. Truncating this expansion at a specific order yields a PDE that approximates the original model. However, the resulting PDEs may not be uniformly valid across the entire domain, especially in the presence of boundary layers or singular regions, where rapidly varying dynamics occur. Moreover, the appropriate order of truncation can be unclear since different choices can lead to inconsistent results across models \cite{lai2025impact,chaplain2020bridging}. Additionally, the number and form of the boundary conditions used to close the PDE depend on the chosen truncation order, and it is not always clear what boundary conditions are appropriate to impose in the continuum model. Similar difficulties arise in agent-based models \cite{chaplain2020bridging,stace2020discrete,kory2024discrete,wattis2020mathematical}, where coarse graining \cite{simpson2024discrete,crossley2025modelling} and mean-field approximations \cite{pillay2017modeling} are often used to derive continuum approximations. These unresolved issues demonstrate the need for alternative, more formal approaches to continuum approximations.

The method of multiple scales applied to discrete systems provides a systematic alternative to Taylor expansions for deriving continuum approximations. It analyses difference equations by introducing a discrete short scale and a continuum long scale and uses matched asymptotic expansions to construct uniformly valid approximations. First introduced to study difference equations \cite{hoppensteadt1977multitime}, it has since been applied to linear and nonlinear difference equations \cite{luongo1996perturbation,maccari1999perturbation,rafei2012solving,van2009multiple,mickens1987periodic}, the discrete Painlevé equation \cite{joshi2015stokes}, nonlinear wave propagation \cite{marathe2006wave}, first integrals for difference equations \cite{rafei2010asymptotic}, functional equations \cite{rafei2014constructing}, and the discrete logistic equation \cite{hall2016multiple}. In mathematical biology, it has been used to derive continuum descriptions of Delta–Notch signalling \cite{o2011multiscale} and to study random walks on periodic lattices \cite{chapman2017effective}. In the former, periodicity emerges naturally from nearest-neighbour interactions, capturing macroscopic organisation without imposing periodicity assumptions \emph{a priori}.

In this paper, we use the method of multiple scales applied to discrete systems, combined with matched asymptotic expansions, to identify where continuum representations of the discrete model are valid and, in these cases, to systematically derive appropriate continuum approximations  of structured models. Unlike an approach based solely on Taylor expansions, the method of multiple scales approach allows one to produce uniformly valid approximations. It also clarifies which truncations are appropriate in different regions of structure space, and naturally determines the appropriate boundary conditions through a discrete boundary layer analysis, overcoming the limitations identified above. \changed{To demonstrate the method in general terms, we start with a broad class of models in which a population is structured into discrete classes.} Structure might represent age, health status (such as exhaustion, substance accumulation), or the ability to acquire resources and reproduce (e.g. phenotype). \changed{Although our analysis is developed independently of a specific application, we later demonstrate its utility using} a lipid-structured model that characterises macrophage-lipid interactions during the early stages of atherosclerosis \cite{chambers2023new,boren2020low,tabas2016macrophage,back2019inflammation}.

The remainder of this paper is organised as follows. \changed{In \Cref{sec:toy problem} we present a paradigm problem to illustrate the commonly used approach to deriving continuum approximations and highlight the issues that can arise.} In \Cref{sec:model_development}, we introduce the general structured mathematical model that we study. We also present numerical simulations with linear functional forms to motivate the model's asymptotic structure and subsequent analysis. We derive the continuum approximation in \Cref{sec:discrete-to-continuum_approximation}, where we show that this requires different approximations in different asymptotic regions of interest. In \Cref{sec:example}, we apply the analysis to the lipid-structured model of early atherosclerosis mentioned above. Finally, in \Cref{sec:conclusion} we discuss our results and draw conclusions from our analysis.

\section{A paradigm problem}\label{sec:toy problem}
To illustrate the commonly used Taylor-expansion approach and how it can fail, we present a simple paradigm problem that admits an exact discrete solution. Specifically, we consider the discrete system
\begin{subequations}\label{eq:toy problem equation}
\begin{align}
x_0&=0,\label{eq:toy problem 0}\\
\frac{\dd x_n}{\dd t}&=N\left[a\left(\frac{x_{n+1}-x_{n-1}}{2}\right)+b\left(x_{n+1}-2x_n+x_{n-1}\right)\right],\quad 1\leq\dvar\leq\lmax-1,\label{eq:toy problem main eq}\\
x_N&=1\label{eq:toy problem N},
\end{align}
\end{subequations}
where \changed{$2b>a>0$ (for reasons made clear below)} and $N\gg1$. We also impose the following initial conditions
\begin{equation}
    x_n(0)=1,\quad 1\leq\dvar\leq\lmax-1.
\end{equation}
The right-hand side of \Cref{eq:toy problem main eq} represents a simple discrete advection-diffusion operator \changed{in the structure variable $\dvar$. We are interested in understanding where in structure space a continuum approximation of \eqref{eq:toy problem equation} is appropriate, and why the classic Taylor expansion approach can fail in certain regions of structure space.} Following the standard Taylor expansion procedure \cite{lai2025impact,chambers2023new}, we introduce a small parameter $\eps$, a continuous structure variable $s$, and a continuous function $x(s,t)$:
\begin{equation}
    \eps=\lmax^{-1}\ll1, \qquad s= \varepsilon n, \qquad x(s,t)\sim x_n(t).
\end{equation}
Under the assumption that $x$ varies smoothly with respect to $s$, one can expand $x_{n\pm1}$ about $s$
\begin{equation}
x(s\pm\varepsilon,t)=\sum_{k=0}^{\infty}\frac{(\pm\varepsilon)^k}{k!}\frac{\pa^k x}{\pa s^k}.\label{eq:toy problem Taylor expansion}
\end{equation}
Substituting the expansion \eqref{eq:toy problem Taylor expansion} into \Cref{eq:toy problem main eq} and collecting terms of the same order yields the infinite-order PDE
\begin{equation}
\frac{\pa x}{\pa t}
=a\sum_{k=0}^{\infty}\frac{\eps^{2k}}{(2k+1)!}\frac{\pa^{2k+1}x}{\pa s^{2k+1}}
+2b\sum_{k=1}^{\infty}\frac{\eps^{2k-1}}{(2k)!}\frac{\pa^{2k}x}{\pa s^{2k}}.\label{eq:toy problem outer PDE}
\end{equation}
Retaining only leading-order terms produces the first-order PDE 
\begin{equation}
    x_t=ax_s,
\end{equation}
which would generally only require one boundary condition in $s$. However, consideration of the discrete boundary conditions \eqref{eq:toy problem 0} and \eqref{eq:toy problem N} might suggest that we should impose two continuum boundary conditions in $s$:
\begin{equation}
    x(0,t)=0, \quad x(1,t)=1.\label{eq:toy problem BCs}
\end{equation}
A common suggestion to resolve this discrepancy is to retain the next-order diffusive correction from \eqref{eq:toy problem outer PDE}, leading to the second-order PDE
\begin{equation}
\label{eq: 2nd order PDE toy}
    x_t=ax_s+\eps bx_{ss}.
\end{equation}
Imposing the boundary conditions \eqref{eq:toy problem BCs} on the second-order PDE \eqref{eq: 2nd order PDE toy} yields the steady-state solution
\begin{equation}
    x^*(s)=\frac{1-\mathrm{e}^{-as/\eps b}}{1-\mathrm{e}^{-a/\eps b}}\sim 1-\mathrm{e}^{-as/\eps b}.\label{eq:toy problem continuum solution}
\end{equation}

For comparison, we now return to the discrete system \eqref{eq:toy problem equation}. The steady-state solution can be obtained exactly, yielding
\begin{equation}
    x^*_n=\frac{1-\left(\frac{2b-a}{2b+a}\right)^{n}}{1-\left(\frac{2b-a}{2b+a}\right)^{N}}=\frac{1-\exp\left[{-s\log \left(\frac{2b+a}{2b-a}\right)/\eps}\right]}{1-\exp\left[{-\log \left(\frac{2b+a}{2b-a}\right)/\eps}\right]}\sim1-\exp\left[-s\log \left(\frac{2b+a}{2b-a}\right)/\eps\right],\label{eq:toy problem discrete solution}
\end{equation}
\changed{where the asymptotic equivalence is in the limit $0 < \eps \ll 1$.} We also note that we can only write the steady-state solution $x^*_n$ as an exponential if $2b>a$.\footnote{
\changed{When $2b < a$, the discrete solution \eqref{eq:toy problem discrete solution} in $\dvar$ shows that the `smooth' exponential representation breaks down and is replaced by an oscillatory solution (highlighting that the continuum approximation in \Cref{eq:toy problem continuum solution} is even less appropriate in this regime). If \eqref{eq:toy problem equation} were a discrete approximation of a continuum equation, this would represent a numerical instability. However, since we start with a discrete system, this is the genuine system behaviour. An adaptation of the methodology we present here could be used to analyse such behaviours.}}
Comparing the steady solution of the full problem \eqref{eq:toy problem discrete solution} with the steady solution of the truncated continuum model \eqref{eq:toy problem continuum solution}, \changed{we see that the solutions are fundamentally different, highlighting that the approach of truncating a Taylor series approximation does not always give the correct result. That is, while both representations are close to $1$ in most of the domain and exhibit a rapid transition to $0$ near $\cvar=0$ via exponential functions, the associated rates of these transitions differ. This means that the solutions will be significantly different to one another when $\cvar = \order(\eps)$. In particular, the continuum approximation \eqref{eq:toy problem continuum solution} predicts an exponential decay rate of $a/( \eps b)$, compared to $\log[(2b+a)/(2b-a)]/\eps$ for the exact discrete solution \eqref{eq:toy problem discrete solution}. These rapid transitions are therefore not equivalent in general. }

The reason why the procedure fails here is that the Taylor expansion underlying the continuum approximation is only valid if $x$ varies sufficiently slowly with respect to the continuum variable $s$. In regions where the solution varies more rapidly, the expansion may break down. To illustrate this, we introduce the inner boundary layer scaling $s=\varepsilon\tilde{s}$ for the PDE \eqref{eq:toy problem outer PDE}, which becomes 
\begin{equation}
\frac{\pa x}{\pa t}
=\frac{1}{\eps}\left(a\sum_{k=0}^{\infty}\frac{1}{(2k+1)!}\frac{\pa^{2k+1}x}{\pa \tilde s^{2k+1}}+2b\sum_{k=1}^{\infty}\frac{1}{(2k)!}\frac{\pa^{2k}x}{\pa \tilde s^{2k}}\right).
\end{equation}
In this inner scaling, all derivatives with respect to $\tilde{s}$ apparently contribute at the same (leading) asymptotic order, so no finite truncation of the series is asymptotically justified. As a result, the continuum PDE obtained by truncating \Cref{eq:toy problem outer PDE} fails to capture the dynamics in regions where $x$ varies rapidly. Specifically, the truncated continuum model is not valid in boundary or interior layers when there is an $\order(1)$ variation in $x$ over an $\order(\eps)$ region in $s$.

\begin{figure}[ht!]
\centering
\includegraphics[width=0.9\textwidth]{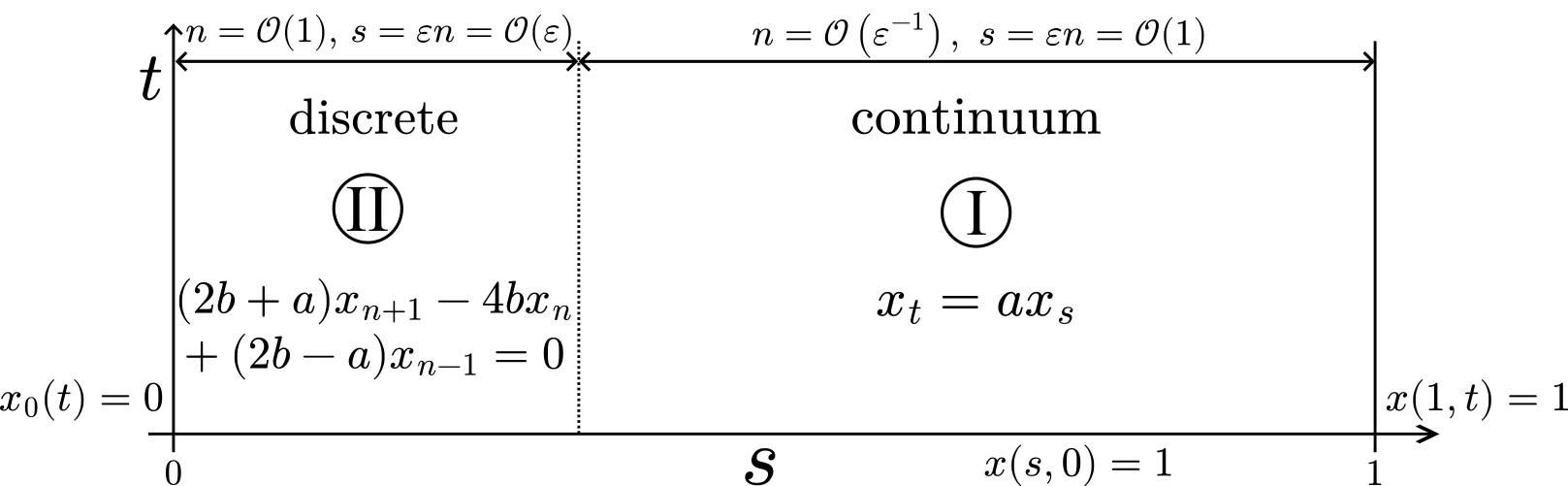}
\caption{The asymptotic structure of the illustrative example consists of one outer region (I) which can be represented with a continuum structure variable, and one boundary layer (II) which is fundamentally discrete. \changed{The size of Region II is exaggerated for illustrative purposes.}}
\label{fig:toy problem asymptotic structure}
\end{figure}

\changed{In general, these issues can be resolved by using the method of multiple scales applied to discrete systems, which we discuss in more detail in \Cref{sec:discrete-to-continuum_approximation}. For the system \eqref{eq:toy problem equation}, the resolution of these issues for $t = \order(1)$ (i.e. neglecting the early-time boundary layer) is that while there are two different asymptotic regions in structure space, only the outer region (Region I in \Cref{fig:toy problem asymptotic structure}) can be represented by a truncated continuum equation. The inner region on the left-hand side of the domain (Region II in \Cref{fig:toy problem asymptotic structure}) is fundamentally discrete, i.e. no truncated continuum approximation of the discrete model holds there.} 

\changed{In more detail, in the continuum outer region (Region I), $n=\order\left(\eps^{-1}\right)$ and the leading-order solution is independent of the discrete variable $\dvar$, so that $x_n(t)\equiv x(s,t)=x(\eps n,t)$.} The dynamics in the outer region are then governed by
\begin{align}
    x_t=ax_s, \qquad x(1,t)=1, \qquad x(s,0)=1,\label{eq:toy ex outer problem}
\end{align}
which is solved by
\begin{equation}
    x(s,t)=1.\label{eq:toy ex outer solution}
\end{equation}
In the fundamentally discrete boundary layer (Region II), $n=\order(1)$ and the leading-order problem is given by
\begin{equation}
    (2b+a)x_{n+1}-4bx_n+(2b-a)x_{n-1}=0, \qquad x_0=0.\label{eq:toy ex inner problem}
\end{equation}
\Cref{eq:toy ex inner problem} is solved by
\begin{equation}
    x_n=A\left[1-\left(\frac{2b-a}{2b+a}\right)^{n}\right].\label{eq:toy ex inner solution}
\end{equation}
\changed{To asymptotically match the outer continuum and inner discrete solutions, we require both solutions to be valid in an intermediate transition region, in a manner related to that for solely continuum problems \cite{van1975perturbation,hall2016multiple}. Given the relative simplicity of this paradigm problem, here it is enough to impose a discrete-to-continuum version of Prandtl's matching condition \cite{van1975perturbation}
\begin{align}
\label{eq: matching toy problem}
\lim_{\cvar \to 0^{+}} x = \lim_{\dvar \to \infty} x_\dvar,
\end{align}
with $x$ and $x_\dvar$ defined in \eqref{eq:toy ex outer solution} and \eqref{eq:toy ex inner solution}, respectively. The left-hand side of \eqref{eq: matching toy problem} is $1$, as can be read off from \eqref{eq:toy ex outer solution}. The right-hand side of \eqref{eq: matching toy problem} is $A$, deduced from the discrete inner solution \eqref{eq:toy ex inner solution} and noting that $2b > a$. Therefore, the matching condition \eqref{eq: matching toy problem} becomes $A = 1$. }
The resulting composite solution
\begin{equation}
    x\sim1-\left(\frac{2b-a}{2b+a}\right)^{n}=1-\left(\frac{2b-a}{2b+a}\right)^{s/\eps}
\end{equation}
is asymptotically consistent with \Cref{eq:toy problem discrete solution} at leading order, demonstrating its accuracy compared to the asymptotically inconsistent solution of the truncated PDE \eqref{eq:toy problem continuum solution}.

\section{\changed{General structured model}}\label{sec:model_development}

\changed{Here, we formulate the model in general terms so that the framework is applicable to a broad class of discretely structured populations. Later, in \Cref{sec:example}, we apply this general framework to a lipid-structured population of macrophages during early atherosclerosis, where transitions between neighbouring structure levels represent lipid uptake and offloading.}

\begin{figure}[ht!]
    \centering
    \includegraphics[width=0.8\linewidth]{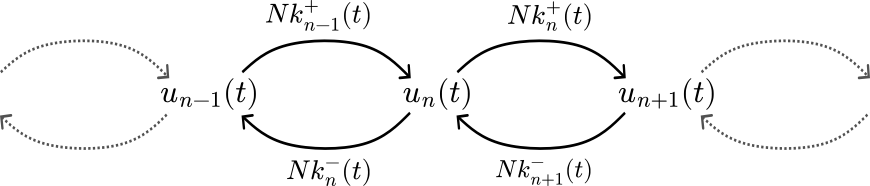}
    \caption{Schematic of forward and backward transitions of neighbouring subpopulations with their respective rates, for $\dvar=1,\dots,\lmax-1$.}
    \label{fig:reactions schematic}
\end{figure}

We consider a population which exhibits an intrinsic organisation (biological structure) and partition it into discrete classes (mathematical structure), presenting the subsequent model in dimensionless form. In this model, $\dfun_{\dvar}(t)\geq0$ is the dimensionless density of the subpopulation with structure level $\dvar\in\{0,1,\dots,\lmax\}$ at time $t$. We consider a model where transitions between neighbouring subpopulations occur at rates $\lmax\klup{\dvar}(t)$ and $\lmax\kldown{\dvar}(t)$ (see \Cref{fig:reactions schematic}).

We impose
\begin{align}
        \klup{\lmax}(t)\equiv0, \qquad \kldown0(t)\equiv0,
        \label{eq:k+-=0}
\end{align}
which ensures that there are no forward transitions by individuals with $\dvar=\lmax$, and no backward transitions by individuals with $\dvar=0$, respectively. We also assume that transitions to higher structure levels become progressively more difficult at higher levels, while transitions to lower levels are increasingly difficult at lower levels, corresponding to $\klup{\dvar}(t)$ and $\kldown{\dvar}(t)$ being monotonically decreasing and increasing in $\dvar$, respectively. 
In this model, we assume further that recruitment is restricted to the endpoints $\dvar=0$ and $\dvar=\lmax$, where it occurs at rates $f(t)$ and $g(t)$, respectively; we discuss the more general case of distributed recruitment in \Cref{sec:conclusion}. Finally, we suppose that individuals with structure level $\dvar$ are removed (via e.g. death or emigration) at rate $\beta_{\dvar}(t)>0$. For $t>0$, our governing equations are
\begin{subequations}\label{eq:dimensional_structured_model}
\begin{align}
    \frac{\mathrm{d}\dfun_0}{\mathrm{d}t}&=\underbrace{f(t)}_{\text{recruitment at }\dvar=0}-\,\lmax\klup0\dfun_0+\lmax\kldown1\dfun_1-\beta_0 \dfun_0,\label{eq:m0_ODE}\\
    \frac{\mathrm{d}\dfun_\dvar}{\mathrm{d}t}&=\underbrace{\lmax\left[\klup{\dvar-1}\dfun_{\dvar-1} - \klup{\dvar}\dfun_\dvar\right]}_{\text{forward transition}}+\underbrace{ \lmax\left[\kldown{\dvar+1}\dfun_{\dvar+1} - \kldown{\dvar}\dfun_\dvar\right]}_{\text{backward transition}} - \underbrace{\beta_{\dvar} \dfun_\dvar}_{\text{removal}}, \quad 1\leq\dvar\leq\lmax-1 \label{eq:ml_ODE}\\
    \frac{\mathrm{d}\dfun_{\lmax}}{\mathrm{d}t}&=\underbrace{g(t)}_{\text{recruitment at }\dvar=\lmax}+\,\lmax\klup{\lmax-1}\dfun_{\lmax-1} -\lmax\kldown{\lmax}\dfun_{\lmax}-\beta_{\lmax} \dfun_{\lmax}.\label{eq:mlmax_ODE}
\end{align}
\end{subequations}

A key assumption for our upscaling is that the functions $\klup{\dvar},\kldown{\dvar},\betal{\dvar}$ are well-defined and continuous in $\dvar\lmax^{-1}$, with bounded gradients as $\lmax\to\infty$. That is, these functions vary slowly in $\dvar$ and, for $\lmax\gg1$, can be approximated by functions of $t$ and
\begin{align}
\cvar:=\dvar\lmax^{-1},\label{eq:definition of \cvar}
\end{align}
so we may write
\begin{equation}\label{eq:ks_independence_of_discrete_variable}
    h_\dvar(t)\equiv h(\cvar,t)=h(\dvar\lmax^{-1},t),
\end{equation}
for $h\in\{\kup,\kdown,\beta\}$.
Then \Cref{eq:k+-=0} gives
\begin{align}
    \kup(1,t)\equiv0, \qquad \kdown(0,t)\equiv0.
    \label{eq:continuous ks=0}
\end{align}
Given continuity, these functions are small close to their vanishing points. That is, we have
\begin{align}
    \kup=o(1)\quad \text{for}\,\,\lmax-\dvar=\order(1), \qquad \kdown=o(1)\quad\text{for}\,\,\dvar=\mathcal{O}(1).\label{eq:k+-small}
\end{align}
We further assume that they are not small away from these points
\begin{equation}
    \kup,\kdown=\order(1),
    \label{eq:ks O(1)}
\end{equation}
and that $\beta,f,g=\order(1)$ within the full structure space. \changed{We define
\begin{equation}\label{eq:K definition}
    K(\cvar,t):=\kup(\cvar,t)-\kdown(\cvar,t).
\end{equation}
For reasons that we will flag later, we also assume that $\kup$ and $\kdown$ are such that $\partial K/\partial \cvar$ is bounded away from zero.}

We close the model \eqref{eq:dimensional_structured_model} by imposing the following initial conditions 
\begin{equation}
    \dfun_{\dvar}(0)=\minitl{\dvar}\geq0,\quad 0\leq\dvar\leq\lmax.\label{eq:discrete ICs}
\end{equation}
We assume that the sum of the initial distribution, $\sum_{\dvar=0}^{\lmax}\minitl{\dvar}$, \changed{is of $\order(1)$ as $\lmax\to\infty$,} implying that the initial mass of the system remains finite, so that refining the discretisation does not alter the total mass. \changed{To impose this, we say that each $\minitl{\dvar}$ scales with $\lmax^{-1}$.} We also assume that \(\minitl{\dvar}\) represents samples from a smooth profile, so that it varies slowly with respect to \(\dvar\). For \(\lmax \gg 1\), it can therefore be approximated by a function of the continuous variable \(\cvar\) (defined in \Cref{eq:definition of \cvar}), namely
\begin{align}
     \minitl{\dvar} \equiv \changed{\lmax^{-1}}\minit(\cvar)= \changed{\lmax^{-1}}\minit\left(\dvar \lmax^{-1}\right).\label{eq:ICs slowly varying}
\end{align}
We make this assumption for simplicity, but note that the analysis can be extended to initial conditions that are non-smooth in \(\cvar\).

\subsection{Numerical simulations: linear neighbouring transition rates}\label{sec:numerical simulations}

To gain some intuition into the behaviour of solutions to \eqref{eq:dimensional_structured_model}, we first present some numerical simulations. \changed{We consider a system with linear functional forms for $\klup{\dvar}$ and $\kldown{\dvar}$, and constant values for $f,g$ and $\betal{\dvar}$, a choice motivated by an application to early atherosclerosis that we explore in more detail in \Cref{sec:example}}:
\begin{equation}\label{eq:linear_functions}%
    \klup{\dvar}(t)=\constup\left(1-\frac{\dvar}{\lmax}\right), \quad \kldown{\dvar}(t)=\constdown\left(\frac{\dvar}{\lmax}\right), \quad
    f(t)=\frac{1}{2}, \quad  g(t)=0, \quad \beta_{\dvar}(t)=\gamma.
\end{equation}
Here, $\constup,\constdown$ and $\gamma$ are positive parameters. We impose uniform initial conditions so that $\dfun_{\dvar}(0)=\lmax^{-1}$ for $0\leq\dvar\leq\lmax$.

\begin{figure}[ht!]
\centering
\includegraphics[width=1\linewidth]{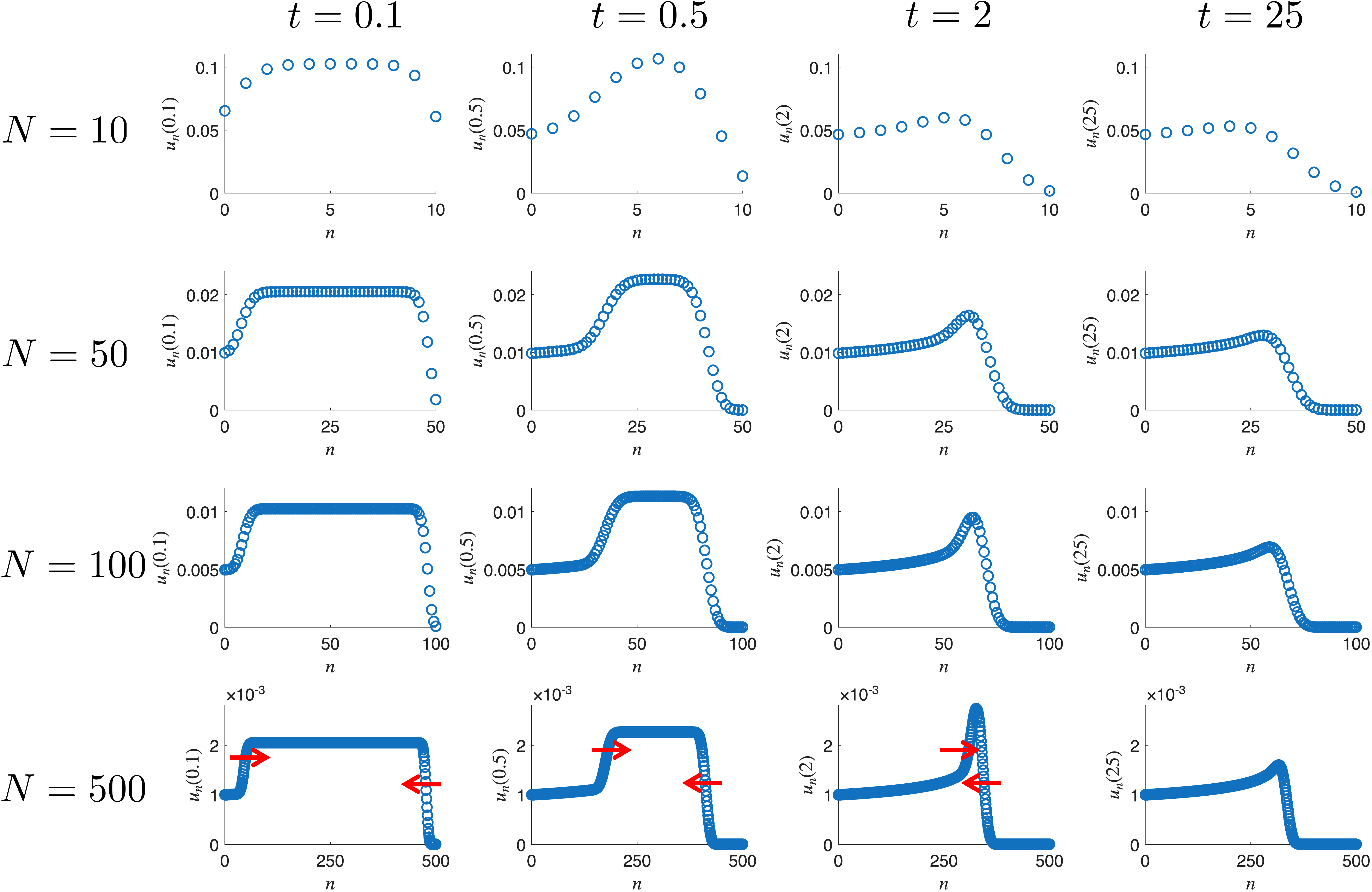}
\caption{Distributions of the discretely structured population, $\dfun_{\dvar}(t)$ for $\dvar=0,1,\dots,\lmax$ found by solving \Cref{eq:dimensional_structured_model,eq:linear_functions} for $\lmax=10,50,100,500$ at times $t=0.1,0.5,2,25$ when $\constup=1,\constdown=0.5,\gamma=1.25$ and $\dfun_{\dvar}(0)=\lmax^{-1}$ for all $\dvar$. For $\lmax=500$, we highlight the apparent moving wavefronts with arrows.}
\label{fig:lmax_increasing}
\end{figure}

We use the ode45 routine in MATLAB with $\lmax=10,50,100$ and $500$ to construct numerical simulations to \Cref{eq:dimensional_structured_model} with functional forms given by \Cref{eq:linear_functions}. The illustrative results in \Cref{fig:lmax_increasing} show the population distribution at early times ($t=0.1$), two intermediate times ($t=0.5,2$) and at longer times ($t=25$). For fixed $\lmax$, the distribution appears to evolve in a wave-like manner. That is, two waves appear to propagate, one from each endpoint in structure space, heading towards an interior point where they coalesce (see arrows for $\lmax=500$). Moreover, we can identify regions of rapid variation throughout, while the distributions appear to be sufficiently continuous to allow a clear interpolation. As $\lmax$ increases, regions of fast variation become sharper, indicating that there may be interior boundary layers for $\lmax\gg1$. %The dynamics at early times ($t=0.1$) reveal a rapid variation in the distribution at the endpoints, suggesting the presence of two boundary layers also at the endpoints. 

\section{Discrete-to-continuum approximation}\label{sec:discrete-to-continuum_approximation}

In this section, we derive a uniformly valid continuum approximation to the general discrete structured model \eqref{eq:dimensional_structured_model} for large $\lmax\gg1$. We transform the discrete distribution $\dfun_{\dvar}(t),\dvar=0,1,\dots,\lmax$, to a continuous function $\dfun_\dvar(\cvar,t)$ defined over a continuous structure variable $\cvar$. We combine the method of multiple scales with matched asymptotic expansions, both applied to discrete systems, as used in for example \cite{hall2016multiple}. The key steps are:
\begin{enumerate}[label=\textbf{\underline{Step~\arabic*}.}, leftmargin=*]
    \item Introduce two length scales: a short discrete scale $\dvar$ and a long continuum scale $\cvar$.
    \item Assume dependence of the solution $\dfun_\dvar(\cvar,t)$ on both length scales. The extra degree of freedom this introduces will be removed later in Step 6, in the usual manner for the standard method of multiple scales.
    \item Self-consistently expand $\dfun_\dvar$ in both the discrete and continuum variables.
    \item Construct an asymptotic expansion for $\dfun_\dvar$ in $\lmax^{-1} \ll 1$.
    \item Solve the leading-order problem and identify a short-scale periodicity.
    \item At higher order, apply the Fredholm alternative theorem with the short-scale periodicity to obtain a solvability condition. This is the requisite upscaled PDE.
\end{enumerate}

In our setting, the discrete class variable $\dvar$ naturally plays the role of the short scale, capturing the fine‐scale variation across neighbouring classes. We introduce a corresponding continuum long scale by defining $\cvar$ as in \Cref{eq:definition of \cvar}, and set $\eps=\lmax^{-1}$ to obtain
\begin{align}
\cvar = \eps \dvar,
\end{align}
considering $\cvar$ to vary smoothly over the interval $[0,1]$ as $\dvar$ ranges from $0$ to \(\lmax\). We then regard the solution as depending simultaneously on both variables.
\changed{In our model, we will find that $\dfun_\dvar(\cvar,t)$ ends up being formally independent of $\dvar$ (or, equivalently, 1-periodic in $\dvar$) at leading order in several asymptotic regions. In the method of multiple scales the additional degree of freedom introduced by initially treating the short and long scales as independent variables is later constrained by periodicity on the short scale. Here, $1$-periodicity in the discrete variable $\dvar$ implies that neighbouring populations are similar over the short scale, so that any significant variation occurs only over the long continuum scale $\cvar$. In general, however, $p$-periodicity can be captured using a similar approach to that presented here. Further details on the discrete multiple scales approach and the role of short-scale periodicity can be found in \cite{hoppensteadt1977multitime,hall2016multiple}.}

\subsection{Asymptotic structure}
\begin{figure}[ht!]
\centering
\includegraphics[width=1\linewidth]{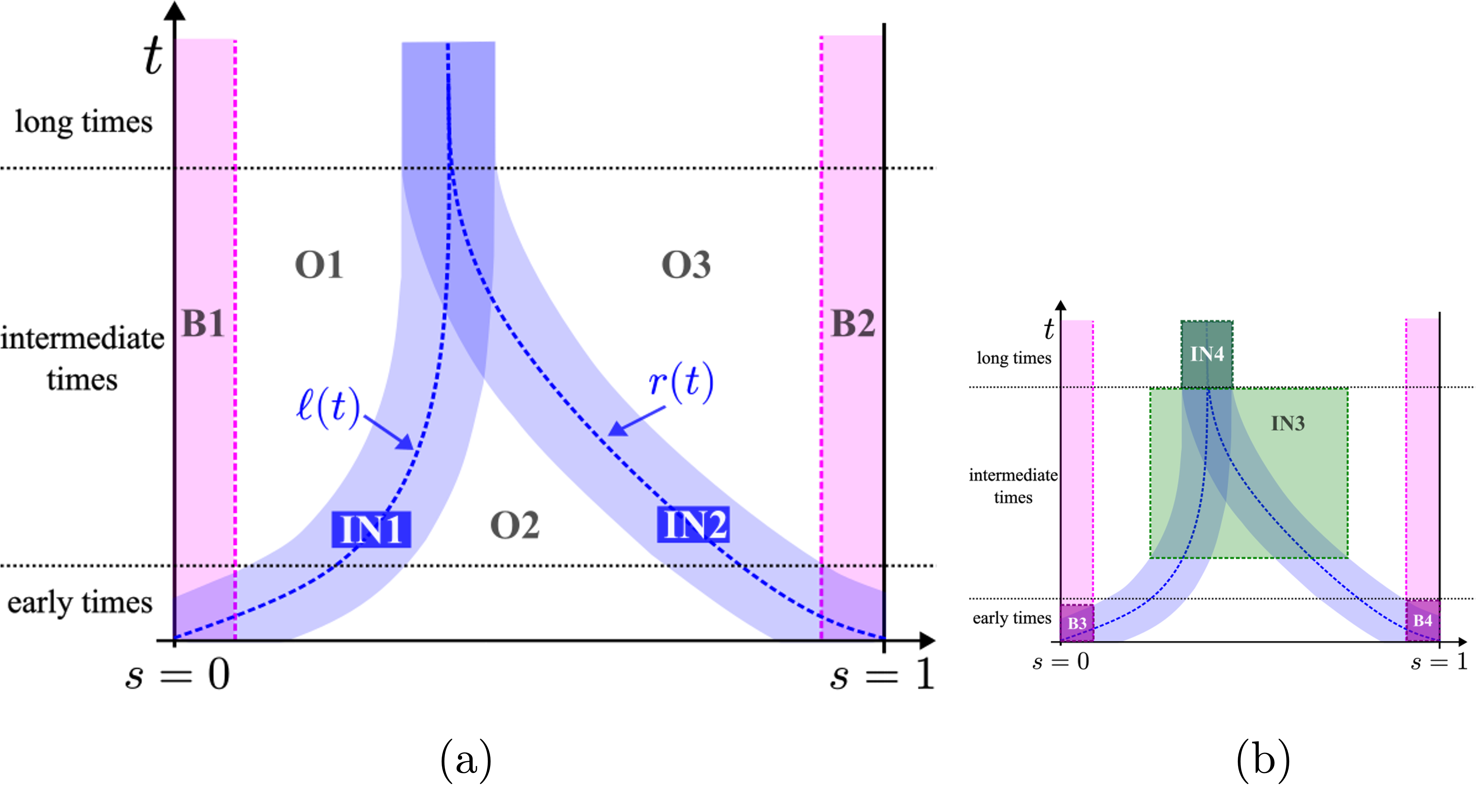}
\caption{{\normalfont(a)} The asymptotic regions we investigate in this paper, via the multiple scales analysis in \Cref{sec:outer regions,sec:left_BL,sec:right_BL,sec:effective BCs,sec:characteristics BL}. The outer regions (white, O1–O3) are governed by continuum equations \eqref{eq:secularity_condition_outer}, while the shaded zones correspond to boundary layers where the dynamics change character. At the domain endpoints, boundary layers B1 and B2 arise; because these regions span only an $\order(1)$ number of structure classes, rather than a large number, they are fundamentally discrete and cannot be described by a continuum approximation. Within the interior, moving fronts at $\cvar=\sstar(t)$ (\Cref{eq:characteristic definition,eq:wavefronts ICs} with $\charac=\sstar$) and $\cvar=\sstarr(t)$ (\Cref{eq:characteristic definition,eq:wavefronts ICs} with $\charac=\sstarr$) generate additional inner layers (regions IN1 and IN2). \changed{The widths of the layers are exaggerated for illustrative purposes.} {\normalfont(b)} Additional asymptotic regions arise due to the intersections of various boundary layers, which we do not investigate in this paper, \changed{with the exception of IN4 in the application considered in \Cref{sec:example}}. At early times, the inner boundary layer at the left wavefront intersects the boundary layer at the left endpoint, forming region B3. Similarly, region B4 arises from the right endpoint-wavefront intersection. As time evolves, the two inner layers intersect to form a transient overlap region (region IN3), and they eventually coalesce into a single steady-state inner layer (region IN4).}
\label{fig:initial_asymptotic_structure}
\end{figure}

Before presenting the technical details of our upscaling, it is helpful to briefly outline the asymptotic structure we identify and subsequently analyse. We illustrate schematically this asymptotic structure in \Cref{fig:initial_asymptotic_structure}. \changed{The solution structure consists of two inner boundary layers (regions IN1 and IN2) which separate three outer regions (regions O1, O2 and O3), with two additional boundary layers at the left and right endpoints (regions B1 and B2). The inner and outer regions are visible in the simulations shown in \Cref{fig:lmax_increasing}.} Additional asymptotic regions will arise when these boundary layers overlap, but for brevity we do not investigate those here. As $t$ increases, the two inner boundary layers start to intersect, forming a transient overlap region (region IN3) that separates outer regions O1 and O3, and which eventually reduces
to a single inner layer (region IN4). Moreover, for early times, the boundary layer at the left endpoint intersects the inner boundary layer at the left wavefront, forming region B3 (and similarly for region B4 on the right-hand side of the domain).

Below, we start by deriving the continuum approximation in the outer regions. Then, we focus on the boundary layers at the domain endpoints to derive appropriate boundary conditions for the continuum problem. Finally, we conduct boundary layer analysis within the inner boundary layers to close the continuum problem.

\subsection{Outer regions}\label{sec:outer regions}
We first consider the outer regions away from the endpoints at $\dvar=0,\lmax$, where $\eps\dvar,1-\eps\dvar=\order(1)$. In these outer regions, the system dynamics are given by \Cref{eq:ml_ODE}. We apply the method of multiple scales to the discrete system by following Steps 1-6 specified at the beginning of \Cref{sec:discrete-to-continuum_approximation}.

\bigskip
\noindent\textbf{\underline{Step 1.}}\\
Using the method of multiple scales applied to discrete systems, we introduce  $\cvar=\eps \dvar\in(0,1)$ with $s=\order(1)$ (defined in \Cref{eq:definition of \cvar}) as the continuum scale.

\bigskip
\noindent\textbf{\underline{Step 2.}}\\
We assume that the solution $\dfun_{\dvar}$ to \Cref{eq:dimensional_structured_model} depends on both the discrete and continuum scales:
\begin{equation}
    \dfun_{\dvar}(t)\equiv \dfun_{\dvar}(\cvar,t)=\dfun_{\dvar}(\varepsilon\dvar,t).
\end{equation}
Moreover, from \Cref{eq:ks_independence_of_discrete_variable} we have
\begin{align}
    \klup{\dvar}(t)\equiv\kup(\cvar,t)=\kup(\eps\dvar,t), \,\, \kldown{\dvar}(t)\equiv\kdown(\cvar,t)=\kdown(\eps\dvar,t), \,\, \betal{\dvar}(t)\equiv\beta(\cvar,t)=\beta(\eps\dvar,t).
\end{align}

\noindent\textbf{\underline{Step 3.}}\\
We self-consistently expand in the discrete and continuum variables:
\begin{subequations}\label{eq:Taylor_expansions_outer_regions}%
\begin{align}
    \dfun_{\dvar\pm1}(\varepsilon\dvar\pm\varepsilon,t)&=\dfun_{\dvar\pm1}(\cvar\pm\varepsilon,t)= \dfun_{\dvar\pm1}(\cvar,t)\pm\varepsilon\frac{\partial \dfun_{\dvar\pm1}}{\partial \cvar}(\cvar,t)+\mathcal{O}(\eps^2),\\
    \kup(\varepsilon\dvar-\varepsilon,t)&=\kup(\cvar-\varepsilon,t)= \kup(\cvar,t)-\varepsilon\frac{\partial \kup}{\partial \cvar}(\cvar,t)+\mathcal{O}(\eps^2),\\
    \kdown(\varepsilon\dvar+\varepsilon,t)&=\kdown(\cvar+\varepsilon,t)= \kdown(\cvar,t)+\varepsilon\frac{\partial \kdown}{\partial \cvar}(\cvar,t)+\mathcal{O}(\eps^2).
\end{align}
\end{subequations}
Substituting \Cref{eq:Taylor_expansions_outer_regions} into \Cref{eq:ml_ODE}, we obtain
\begin{align}
    \frac{\partial \dfun_{\dvar}}{\partial t}&=  \frac{1}{\eps}\left[\left(\kup-\varepsilon\frac{\partial \kup}{\partial \cvar}\right)\left(\dfun_{\dvar-1}-\varepsilon\frac{\partial \dfun_{\dvar-1}}{\partial \cvar}\right)-\kup \dfun_{\dvar} \right]\nonumber\\
    &+\frac{1}{\eps}\left[\left(\kdown+\varepsilon\frac{\partial \kdown}{\partial \cvar}\right)\left(\dfun_{\dvar+1}+\varepsilon\frac{\partial \dfun_{\dvar+1}}{\partial \cvar}\right) -\kdown \dfun_{\dvar}\right]-\beta \dfun_{\dvar}+\mathcal{O}(\eps).
    \label{eq:multiple_scales_only_expansion}
\end{align}
In \Cref{eq:multiple_scales_only_expansion} we have omitted the function arguments for notational brevity but here, and hereafter, $\dfun_{\dvar},\kupdown$ and $\beta$ are evaluated at $(\cvar,t)$ unless otherwise specified.

\bigskip
\noindent\textbf{\underline{Step 4.}}\\
We expand $\dfun_{\dvar}(\cvar,t)$ as an asymptotic series: 
\begin{equation}
    \dfun_{\dvar}(\cvar,t)=\eps\ml0(\cvar,t)+\eps^2\ml1(\cvar,t)+\mathcal{O}(\eps^3),\label{eq:asymptotic expansion outer regions}
\end{equation}
where $\ml i=\order(1)$.
\changed{The leading-order scaling in \Cref{eq:asymptotic expansion outer regions} ensures that the mass of the system is independent of $\lmax=\eps^{-1}$ as $\lmax\to\infty$, so that it remains finite and $\order(1)$. This is because the mass of the system is distributed across $\lmax=\varepsilon^{-1}$ discrete structure classes, therefore the mass associated with each class must scale as $\order(\lmax^{-1})=\order(\varepsilon)$.}
Substituting \Cref{eq:asymptotic expansion outer regions} into \Cref{eq:multiple_scales_only_expansion}, we obtain
\begin{align}\label{eq:outer regions full expansion PDE}
    \eps\frac{\partial \dfun_{\dvar}^{(0)}}{\partial t}= &\left[\kup\left(\dfun_{\dvar-1}^{(0)}-\dfun_{\dvar}^{(0)}\right)+\kdown\left(\dfun_{\dvar+1}^{(0)}-\dfun_{\dvar}^{(0)}\right) \right] \nonumber\\
    &+\eps \left[\kup\left(-\frac{\partial \dfun_{\dvar-1}^{(0)}}{\partial \cvar}+\dfun_{\dvar-1}^{(1)}-\dfun_{\dvar}^{(1)}\right)-\frac{\partial\kup}{\pa \cvar}\dfun_{\dvar-1}^{(0)}\right.\nonumber\\
    &\left. +\kdown\left(\frac{\partial \dfun_{\dvar+1}^{(0)}}{\partial \cvar}+\dfun_{\dvar+1}^{(1)}-\dfun_{\dvar}^{(1)}\right)+\frac{\pa\kdown}{\pa \cvar}\dfun_{\dvar+1}^{(0)}-\beta \dfun_{\dvar}^{(0)}\right]+\mathcal{O}(\eps^2).
\end{align}

\bigskip
\noindent\textbf{\underline{Step 5.}}\\
Equating the $\mathcal{O}\left(1\right)$ terms in \Cref{eq:outer regions full expansion PDE} yields
\begin{align}
    0= \kup\left(\dfun_{\dvar-1}^{(0)}-\dfun_{\dvar}^{(0)}\right)+\kdown\left(\dfun_{\dvar+1}^{(0)}-\dfun_{\dvar}^{(0)}\right).\label{eq:previous_eq}
\end{align}
The general solution to the linear discrete equation \eqref{eq:previous_eq} is
\begin{equation}
   \dfun_{\dvar}^{(0)}(\cvar,t)=A(\cvar,t)+B(s,t)\left(\frac{\kup}{\kdown}\right)^\dvar,
\label{eq:outer_region_leading_order_equality}
\end{equation}
recalling that $\dvar=\order(\eps^{-1})$. \changed{The terms that vary exponentially over the short scale $\dvar$ are not asymptotically self-consistent to include here (as corroborated by the numerical solutions).}\footnote{
\changed{This is similar to how exponentially growing solutions in boundary layers are often inconsistent within standard singular ODE problems.}}
Therefore, we have $B(s,t)=0$ and
\begin{align}
    \dfun_{\dvar}^{(0)}(\cvar,t)=A(s,t):=\mouter(\cvar,t),\label{eq:outer regions leading order solution}
\end{align}
where we redefine $A$ as $\mouter$ for notational clarity. We conclude that the solution is independent of the discrete variable $\dvar$ at leading order, and so in these outer regions any variation in the structure variable occurs over the long continuum scale $\cvar$. \changed{Importantly for the method of multiple scales, this formal independence of $\dvar$ means that the leading-order solution \eqref{eq:outer regions leading order solution} is 1-periodic in the short scale $\dvar$, and therefore that all structure variation can be represented solely in terms of the continuum variable $\cvar$ in the outer regions.} Our remaining task is to understand how the leading-order solution $\dfun(\cvar,t)$ depends on the long scale $\cvar$ and time $t$. To do this, we must proceed to higher orders.

\bigskip
\noindent\textbf{\underline{Step 6.}}\\
At $\order(\eps)$ in \Cref{eq:outer regions full expansion PDE}, we have
\begin{align}
    \frac{\partial \mouter}{\partial t}= \frac{\partial}{\partial \cvar}\left[\left(\kdown-\kup\right)\mouter\right]-\beta\mouter+\kup\left(\dfun_{\dvar-1}^{(1)}-\dfun_{\dvar}^{(1)}\right)+\kdown\left(\dfun_{\dvar+1}^{(1)}-\dfun_{\dvar}^{(1)}\right).\label{eq:outer_region_order1_eq}
\end{align}
Imposing the method of multiple scales constraint of 1-periodicity on the short scale (which, together with the leading-order periodicity, maintains uniform validity of the asymptotic expansion) and applying the Fredholm alternative theorem for linear difference operators (see Appendix~\ref{app:FAT}) yields the solvability condition
\begin{align}
    \frac{\partial \mouter}{\partial t}+\frac{\partial}{\partial \cvar}\left(K\mouter\right)=-\beta\mouter,
\label{eq:secularity_condition_outer}
\end{align}
where $K(\cvar,t)$ is defined in \Cref{eq:K definition}. We present a more detailed derivation of the appropriate solvability condition in terms of suitable inner products and adjoints in Appendix \ref{app:FAT}.

The first-order hyperbolic PDE \eqref{eq:secularity_condition_outer} governs the leading-order dynamics of the structured population in the outer regions. To close \Cref{eq:secularity_condition_outer}, we require appropriate boundary and initial conditions. The initial conditions are
\begin{align}
\mouter(\cvar,0)=\minit(\cvar),
\end{align}
where $\minit(\cvar)$ is given by \Cref{eq:discrete ICs,eq:ICs slowly varying}.

\begin{figure}
    \centering
    \includegraphics[width=0.4\linewidth]{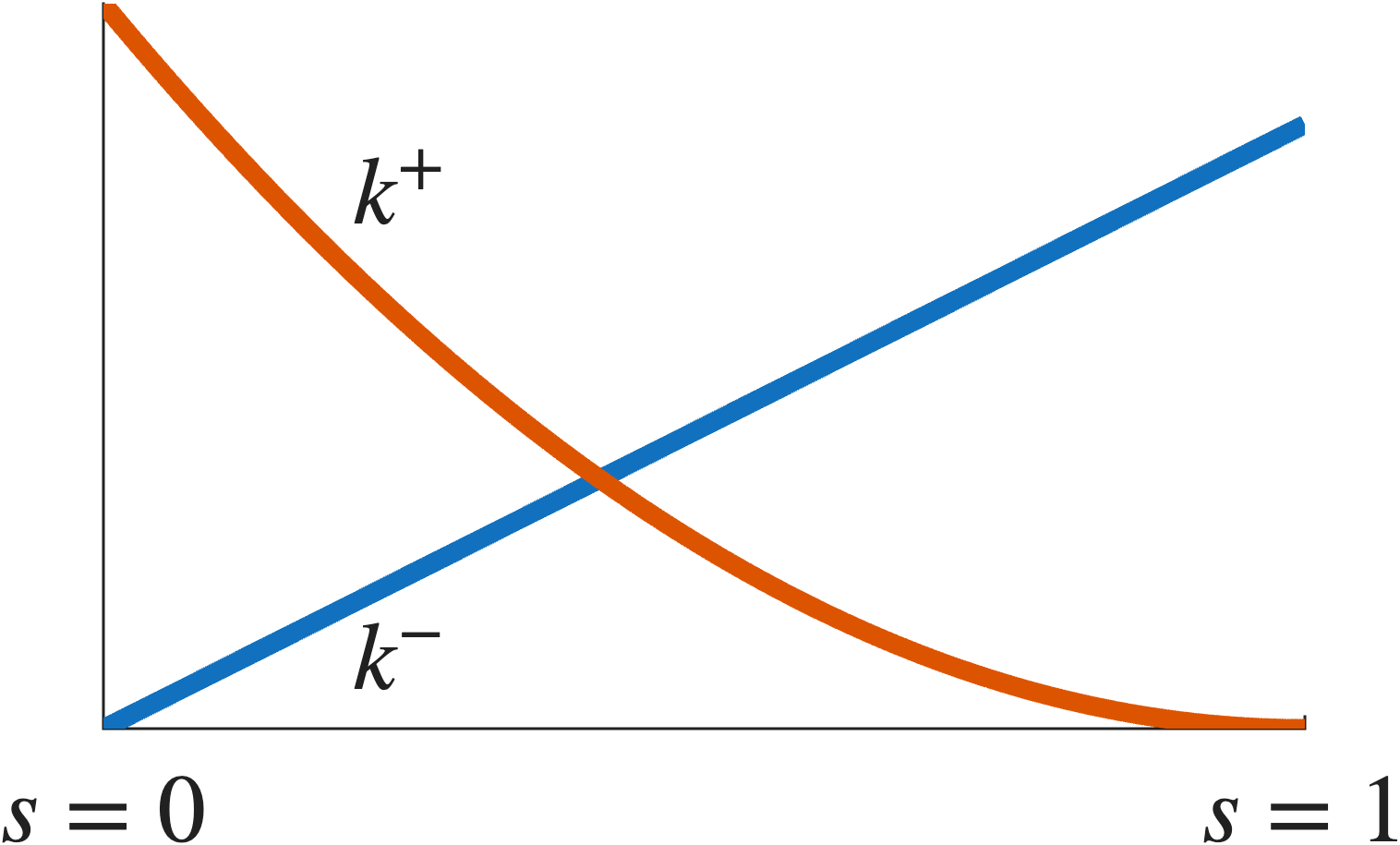}
    \caption{Illustrative sketch showing $\kup$ and $\kdown$ against $\cvar$. Monotonicity with respect to $\dvar$ \changed{and the assumption that $\pa K/\pa \cvar<0$} imply that $\kup$ and $\kdown$ must intersect at a unique interior point in the continuum structure space. At this point, the advection term in \Cref{eq:secularity_condition_outer} is zero and \changed{the steady-state outer equation becomes singular}.}
    \label{fig:k+- sketch}
\end{figure}

At this stage, we do not have enough information to determine the boundary conditions for \Cref{eq:secularity_condition_outer}. Given that \Cref{eq:secularity_condition_outer} is a first-order PDE, we might expect that one boundary condition would suffice. However, information propagates along the characteristics that emerge from the left boundary ($\cvar=0$) to the right, and vice versa from the right boundary ($\cvar=1$). \Cref{fig:k+- sketch} illustrates the transition rates $\kup$ and $\kdown$ as functions of the continuum variable $\cvar$, highlighting how they determine the direction in which information propagates along the characteristics, given by the sign of $K(s,t)$. Therefore two boundary conditions are needed: one associated with the region influenced by characteristics originating at the left boundary, the other with the region influenced by characteristics originating at the right. We derive the appropriate boundary conditions by analysing additional boundary layers at the endpoints of the domain. These asymptotic regions are fundamentally discrete, in that continuum representations are not appropriate therein, as we will show in \Cref{sec:left_BL}. Nevertheless, by matching the discrete problems in these boundary layers to the continuum outer regions, we can derive consistent boundary conditions for the outer problems.

We also note that the coefficient of the advective term in \Cref{eq:secularity_condition_outer} vanishes in the interior of the domain, at the unique location where $K(s,t)=0$ (see \Cref{fig:k+- sketch}), towards which the characteristics are directed.

\subsection{Boundary layer at the left endpoint}\label{sec:left_BL}
We first investigate the asymptotic region B1 near $\dvar=0$. \changed{Our aim is to perform an asymptotic matching on the B1 solution we deduce, in order to derive the appropriate matching condition (and hence an effective boundary condition) for the outer region O1.} A classic Taylor expansion method might suggest a natural approach for investigating the boundary layer in the continuum variable $\cvar$, as for standard PDEs. However, the appropriate scaling would be $\cvar=\order(\eps)$, which corresponds to $\dvar=\order(1)$. As such, it is not consistent to treat $s = \eps n$ as a continuum variable in this region. In fact, the system is fundamentally discrete in this region, where it is governed by \Cref{eq:ml_ODE} with $\dvar=\mathcal{O}(1)$. However, since the coefficients still vary slowly, and we are effectively zooming into a short scale in this region, we may self-consistently expand \Cref{eq:ks_independence_of_discrete_variable} as follows
\begin{subequations}\label{eq:ks Taylor exp in left endpoint BL}%
\begin{align}
    \kupdown_{\dvar}(t)&\equiv \kupdown(\eps\dvar,t)=\kupdown(0,t)+\eps\dvar\frac{\pa\kupdown}{\pa \cvar}(0,t)+\order(\eps^2),\\
    \klup{\dvar-1}(t)&\equiv\kup(\eps\dvar-\eps,t)=\kup(0,t)+(\eps\dvar-\eps)\frac{\pa\kup}{\pa \cvar}(0,t)+\order(\eps^2),
\end{align}
\end{subequations}
with the equivalent for $\kldown{\dvar+1}$ and $\betal{\dvar}$.
Additionally, in this boundary layer we still expand $\dfun_{\dvar}(t)$ as an asymptotic series:
\begin{align}
    \dfun_{\dvar}(t)=\eps\ml0(t)+\eps^2\ml1(t)+\mathcal{O}(\eps^3).\label{eq:left_BL_asymptotic_expansion}
\end{align}
Substituting Equations \eqref{eq:continuous ks=0}, \eqref{eq:ks Taylor exp in left endpoint BL}, \eqref{eq:left_BL_asymptotic_expansion} into \Cref{eq:ml_ODE}, we obtain for $\dvar \geq 1$
\begin{align}
    \eps\frac{\dd\ml0}{\dd t}&= \frac{1}{\eps}\left[\left(\kup(0)+(\eps\dvar-\eps)\frac{\pa\kup}{\pa \cvar}(0)\right)\left(\eps\mlm0+\eps^2\mlm1\right)\right.\nonumber\\
    &\quad\quad\quad-\left.\left(\kup(0)+
    \eps\dvar\frac{\pa\kup}{\pa \cvar}(0)\right)\left(\eps\ml0+\eps^2\ml1\right) \right]\nonumber\\
    &\quad+\frac{1}{\eps}\left[\left(\eps\dvar+\eps\right)\frac{\pa\kdown}{\pa \cvar}(0)\left(\eps\mlp0+\eps^2\mlp1\right)-\eps\dvar\frac{\pa\kdown}{\pa \cvar}(0)\left(\eps\ml0+\eps^2\ml1\right) \right]\nonumber\\
    &\quad-\left(\beta(0)+\eps\dvar\frac{\pa\beta}{\pa \cvar}(0)\right)\left(\eps\ml0+\eps^2\ml1\right)+\order(\eps^2),
\end{align}
where we have omitted explicit dependence on $t$ for notational brevity. Since $\kup,\kdown$ have bounded gradients as $\eps\to0$, we have $\frac{\pa\kupdown}{\pa \cvar}(0)=\order(1)$.
Thus, at $\mathcal{O}(1)$, for $t>0$,
\begin{equation}
    0=\kup(0,t)\left[\mlm0(t)-\ml0(t)\right],
\end{equation}
which gives for $\dvar\geq1$
\begin{equation}
    \mlm0(t)=\ml0(t)=:\dfun_{\text{left}}(t).\label{eq:left_boundary_leading_order_equality}
\end{equation}
In particular, for $\dvar=1$ this gives $\dfun^{(0)}_0(t)=\dfun^{(0)}_1(t)=\dfun_{\text{left}}(t)$.

\Cref{eq:m0_ODE} defines the dynamics of $\dfun_0$, which, given the asymptotic expansion \eqref{eq:left_BL_asymptotic_expansion} and \Cref{eq:k+-=0}, becomes 
\begin{align}
\eps\frac{\mathrm{d}\dfun^{(0)}_0}{\mathrm{d}t}=f(t)-\klup0 \left( \dfun^{(0)}_0+\eps \dfun^{(1)}_0\right)+\kldown1\left( \dfun_1^{(0)}+\eps \dfun_1^{(1)}\right)-\eps\beta_0 \dfun^{(0)}_0+\mathcal{O}(\eps^2).\label{eq:left BL zero equation}
 \end{align}
Equations \eqref{eq:k+-small}-\eqref{eq:ks O(1)} imply that $\klup0=\order(1),\kldown1=o(1)$. Hence using \Cref{eq:left_boundary_leading_order_equality} in \Cref{eq:left BL zero equation} yields at leading order
\begin{align}
    \dfun_{\text{left}}(t)=\frac{f(t)}{\klup0(t)}.\label{eq:left BL solution}
\end{align}
\changed{Using the method of matched asymptotic expansions, the solution \eqref{eq:left BL solution} can be used to obtain the effective boundary condition for the PDE \eqref{eq:secularity_condition_outer} as $\cvar \to 0^{+}$. Specifically, we match the continuum solution $\mouter(\cvar,t)$ as $\cvar\to0^+$ with the far-field limit of the discrete boundary-layer solution \eqref{eq:left BL solution} as $\dvar\to\infty$, giving the matching condition
\begin{align}\label{eq:matching condition}
    \lim_{\cvar\to0^+}\mouter(\cvar,t)= \lim_{\dvar\to\infty}\dfun_{\dvar}^{(0)}(t)=\dfun_{\text{left}}(t).
\end{align}
Thus, \Cref{eq:matching condition} yields the effective boundary condition}
\begin{align}
    \dfun(0^+,t)=\frac{f(t)}{\klup0(t)}.\label{eq:left_BC}
\end{align}
\changed{Physically, this effective boundary condition represents a leading-order balance between recruitment into the lowest structural class and transitions out of that class, as can be seen by writing \eqref{eq:left_BC} as $f(t)=\klup{0}(t)\mouter(0,t)$.}

\subsection{Boundary layer at the right endpoint}\label{sec:right_BL}
In a similar manner to \Cref{sec:left_BL}, the region near the right endpoint $\dvar=\lmax$ is fundamentally discrete and cannot be approximated by a truncated continuum representation. We consider this region by setting $\dvar=\lmax-\hat\dvar=\eps^{-1}-\hat\dvar$ where $\hat\dvar=\mathcal{O}(1)$. The analysis follows that in \Cref{sec:left_BL}, resulting in the leading-order solution
\begin{equation}
    \dfun_{\text{right}}(t):=\dfun^{(0)}_{\lmax}(t)=\frac{g(t)}{\kldown{\lmax}(t)}.\label{eq:m_lmax_expression}
\end{equation}
Matching with the outer solution as $\cvar\to1^-$ gives the effective boundary condition
\begin{equation}
    \dfun(1^-,t)=\frac{g(t)}{\kldown{\lmax}(t)}.\label{eq:right_BC}
\end{equation}

\subsection{Effective boundary conditions for the outer PDE}\label{sec:effective BCs}

Taken together, \Cref{eq:left_BC,eq:right_BC} define two boundary conditions for \Cref{eq:secularity_condition_outer} at $\cvar=0,1$. In the left outer region (region O1, in \Cref{fig:initial_asymptotic_structure}a), characteristics propagate rightward from $\cvar=0$. In the right outer region (region O3, in \Cref{fig:initial_asymptotic_structure}a), they propagate leftward from $\cvar=1$. In region O2 the characteristics emanate from the initial conditions. \changed{The characteristic projections from the outer PDE \eqref{eq:secularity_condition_outer} are directed towards the instantaneous stagnation point, where $K(\cvar,t)=0$.} If there is a discontinuity between the boundary condition at $\cvar=0$ and the initial condition for $\dvar=0$, i.e. if
\begin{equation}
    \frac{f(0)}{\klup{0}(0)}\neq \eps^{-1}\minitl0=\minit(0),
    \label{eq:left discontinuity}
\end{equation}
then the discontinuity propagates along the characteristic projection passing through $(\cvar,t)=(0,0)$, which we identify as the left wavefront (region IN1, in \Cref{fig:initial_asymptotic_structure}a). \changed{Given this apparent discontinuity at $t=0$, there is also an early-time boundary layer, which we do not consider here.
As with the case for $\cvar=0$, any discontinuity between the boundary condition at $\cvar=1$ and the initial condition for $\dvar=\lmax$ $\left(\frac{g(0)}{\kldown{\lmax}(0)}\neq \eps^{-1}\minitl{\lmax}=\minit(1)\right)$ propagates along the characteristic projection passing through $(\cvar,t)=(1,0)$ (right wavefront, region IN2 in \Cref{fig:initial_asymptotic_structure}a). We investigate the dynamics in the wavefronts by performing an inner boundary layer analysis.}

\subsection{Resolving discontinuities across characteristics}\label{sec:characteristics BL}

\changed{Discontinuities in the outer solution propagate along characteristics of the governing PDE. To investigate the dynamics in the regions of rapid variation surrounding these discontinuities, we perform a boundary layer analysis centred on a general characteristic, before identifying the characteristics relevant to our system. Let $\charac(t)$ be a characteristic of \Cref{eq:secularity_condition_outer}:
\begin{align}\label{eq:characteristic definition}
    \dot\charac=K(\charac,t)=\kup(\charac,t)-\kdown(\charac,t),
\end{align}
where the overdot represents differentiation with respect to time.
The appropriate scaling for the interior layer around the characteristic at $\cvar=\charac(t)$ is (for reasons made clear below)
\begin{equation}
    \dvar=\frac{\charac(t)}{\eps}+\frac{\tilde \cvar}{\sqrt{\eps}}\quad\Rightarrow\quad \tilde \cvar=\sqrt{\eps}\left(\dvar-\frac{\charac(t)}{\eps}\right),
    \label{eq:inner_variables_definition_characteristic}
\end{equation}
where $\dvar-\charac/\eps=\order(1/\sqrteps)$, so that $\tilde \cvar=\order(1)$ is the long scale (continuum variable) in this inner region, and $\charac(t)=\mathcal{O}(1)$.}

Assuming dependence of the solution on both the discrete and continuum scales, as per Step 2 of the method of multiple scales applied to discrete systems, gives
\begin{equation}
    \dfun_{\dvar}(t)\equiv \dfun_{\dvar}(\tilde \cvar,t)=\dfun_{\dvar}\left(\sqrt{\eps}\left(\dvar-\frac{\charac}{\eps}\right),t\right).
\end{equation}
Hence the short variable $\dvar\rightarrow\dvar+1$ implies $\tilde \cvar\rightarrow\tilde \cvar+\sqrt{\eps}$. The time derivative is transformed using \Cref{eq:inner_variables_definition_characteristic} and the chain rule implies
\begin{equation}
    \frac{\pa}{\pa t}\rightarrow\frac{\pa}{\pa t}-\frac{\dot\charac}{\sqrt{\eps}}\frac{\pa}{\pa\tilde \cvar}.
\end{equation}
Self-consistently expanding $\dfun_{\dvar\pm1}(t)$ and using a Taylor series expansion in $\tilde \cvar$, we obtain
\begin{equation}
    \dfun_{\dvar\pm1}(t)=\dfun_{\dvar\pm1}(\tilde \cvar\pm\sqrt{\eps},t)= \dfun_{\dvar\pm1}(\tilde \cvar,t)\pm\sqrt{\eps}\frac{\partial \dfun_{\dvar\pm1}}{\partial \tilde \cvar}(\tilde \cvar,t)+\frac{\varepsilon}{2}\frac{\partial^2\dfun_{\dvar\pm1}}{\partial \tilde \cvar^2}(\tilde \cvar,t)+\mathcal{O}\left(\eps^{3/2}\right).
    \label{eq:Taylor_expansion_zero_advection_BL}
\end{equation}

Recall from \Cref{eq:ks_independence_of_discrete_variable} that $\klupdown{\dvar},\betal{\dvar}$ vary slowly in the discrete variable, and, as such, are functions of $(\eps\dvar,t)$ alone (i.e. they are independent of the short scale $\dvar$ under the method of multiple scales framework). Thus, self-consistently expanding using \Cref{eq:inner_variables_definition_characteristic} and a Taylor series expansion yields
\begin{subequations}    \label{eq:ks_expansions_zero_advection_BL}%
    \begin{align}
        \klupdown{\dvar}\equiv\kupdown(\eps\dvar)=\kupdown\left(\charac+\sqrt{\eps}\tilde \cvar\right)&=\kupdownstar+\sqrt{\eps}\,\tilde \cvar\,\pakupdownstar1+\frac{\eps\tilde \cvar^2}{2}\pakupdownstar2+\order\left(\eps^{3/2}\right),\\
        \klup{\dvar-1}\equiv\kup(\eps\dvar-\eps)&=\kup\left(\charac+\sqrt{\eps}\,\tilde \cvar-\eps\right)\nonumber\\
        &=\kupstar+\left(\sqrt{\eps}\,\tilde \cvar-\eps\right)\pakupstar1+\frac{\eps\tilde \cvar^2}{2}\pakupstar2+\order\left(\eps^{3/2}\right),\\
        \kldown{\dvar+1}\equiv\kdown(\eps\dvar+\eps)&=\kdown\left(\charac+\sqrt{\eps}\,\tilde \cvar+\eps\right)\nonumber\\
        &=\kdownstar+\left(\sqrt{\eps}\,\tilde \cvar+\eps\right)\pakdownstar1+\frac{\eps\tilde \cvar^2}{2}\pakdownstar2+\order\left(\eps^{3/2}\right),\\
        \betal{\dvar}\equiv\beta(\eps\dvar)=\beta\left(\charac+\sqrt{\eps}\tilde \cvar\right)&=\beta_c+\order\left(\sqrt{\eps}\right),
    \end{align}
\end{subequations}
where we omit the time dependence in the arguments for notational brevity. Here, we define
\begin{align}
    \kupdown_c:=\kupdown(\charac( t),t), && \pa^p\kupdown_c:=\frac{\pa^p\kupdown}{\pa \cvar^p}(\charac( t),t)\,\,\text{for}\,\,p=1,2, && \beta_c:=\beta(\charac( t),t).
\end{align}
Moreover, since $\kup,\kdown$ have bounded gradients in the limit $\eps\to0$, we have $\pa^p\kupdown_c=\order(1)$.

Substituting \Cref{eq:Taylor_expansion_zero_advection_BL,eq:ks_expansions_zero_advection_BL} into \Cref{eq:ml_ODE}, we obtain
\begin{align}
    \frac{\pa \dfun_{\dvar}}{\pa t}-\frac{\dot\charac}{\sqrt{\eps}}\frac{\pa \dfun_{\dvar}}{\pa \tilde \cvar}&=\frac{1}{\eps}\left[\left(\kupstar+(\sqrt{\eps}\,\tilde \cvar-\eps)\pakupstar1+\frac{\eps\tilde \cvar^2}{2}\pakupstar2\right)\left(\dfun_{\dvar-1}-\sqrt{\eps}\frac{\pa \dfun_{\dvar-1}}{\pa \tilde \cvar}\right.\right.\nonumber\\
    &\quad\quad\quad+\left.\left.\frac{\eps}{2}\frac{\pa^2 \dfun_{\dvar-1}}{\pa \tilde \cvar^2} \right)
    -\left(\kupstar+\sqrt{\eps}\,\tilde \cvar\,\pakupstar1+\frac{\eps\tilde \cvar^2}{2}\pakupstar2 \right)\dfun_{\dvar}\right]\nonumber\\
    &+\frac{1}{\eps}\left[\left(\kdownstar+(\sqrt{\eps}\,\tilde \cvar+\eps)\pakdownstar1+\frac{\eps\tilde \cvar^2}{2}\pakdownstar2 \right)\left(\dfun_{\dvar+1}+\sqrt{\eps}\frac{\pa \dfun_{\dvar+1}}{\pa \tilde \cvar}\right.\right.\nonumber\\
    &\quad\quad\quad+\left.\frac{\eps}{2}\frac{\pa^2 \dfun_{\dvar+1}}{\pa \tilde \cvar^2} \right)
    -\left.\left(\kdownstar+\sqrt{\eps}\,\tilde \cvar\,\pakdownstar1+\frac{\eps\tilde \cvar^2}{2}\pakdownstar2 \right)\dfun_{\dvar} \right]\nonumber\\
    &-\beta_c \dfun_{\dvar}+\order\left(\sqrt{\eps}\right).
    \label{eq:expansion in zero advection BL PDE}
\end{align}
We expand $\dfun_{\dvar}(\tilde \cvar,t)$ as the asymptotic expansion
\begin{equation}
\changed{
    \dfun_{\dvar}(\tilde \cvar,t)=\eps\dfun_{\dvar}^{(0)}(\tilde \cvar,t)+\eps^{3/2}\,\dfun_{\dvar}^{(1/2)}(\tilde \cvar,t)+\eps^2 \,\ml{1}(\tilde \cvar,t)+\mathcal{O}(\varepsilon^{5/2})}.
    \label{eq:asymptotic_expansion_zero_advection_BL}
\end{equation}
Substituting \Cref{eq:asymptotic_expansion_zero_advection_BL} into \Cref{eq:expansion in zero advection BL PDE}, at $\mathcal{O}(1)$ we obtain
\begin{equation}
     0= \kupstar(t)\left[\dfun_{\dvar-1}^{(0)}(\tilde \cvar,t)-\dfun_{\dvar}^{(0)}(\tilde \cvar,t)\right]+\kdownstar(t)\left[\dfun_{\dvar+1}^{(0)}(\tilde \cvar,t)-\dfun_{\dvar}^{(0)}(\tilde \cvar,t)\right].\label{eq:stagnation BL leading order eq}
\end{equation}
For $\kupstar\neq\kdownstar$, the general solution to the linear discrete equation \eqref{eq:stagnation BL leading order eq} is
\begin{equation}
    \dfun_{\dvar}^{(0)}(\tilde \cvar,t)=\tilde A(\tilde \cvar,t)+\tilde B(\tilde \cvar,t)\left(\frac{\kupstar}{\kdownstar}\right)^\dvar.\label{eq:stagnantion BL leading order solution}
\end{equation}
\changed{As discussed in \Cref{sec:outer regions},
the exponentially varying terms over the short scale $\dvar$
are not asymptotically self-consistent to include in the leading-order solution
here. }%\footnote{\changed{Indeed, in the case where $\kup=\kdown$, the solution \eqref{eq:stagnantion BL leading order solution} allows for linear growth in $\dvar$ across the inner region, which would correspond to a change in asymptotic magnitude across the boundary layer that is not consistent with the outer equations (see \Cref{eq:outer_region_leading_order_equality}) and boundary conditions.}}
Hence, the solution in this inner region is independent of the discrete variable $\dvar$ at leading order.
Thus, we require $\tilde B(\tilde s,t)=0$ and
\begin{equation}
    \dfun_{\dvar}^{(0)}(\tilde \cvar,t)=\tilde A(\tilde s,t):=\madv(\tilde \cvar,t),
\end{equation}
where we redefine $\tilde A$ as $\madv$ for later notational clarity. At the next order, equating the $\order(\sqrteps)$ terms in \Cref{eq:expansion in zero advection BL PDE} yields
\begin{align}
    -\dot\charac\frac{\pa \madv}{\pa \tilde \cvar}=&\left[\kdownstar-\kupstar\right]\frac{\pa \madv}{\pa\tilde \cvar}+\kupstar\left[\mlm{1/2}-\ml{1/2}\right]+\kdownstar\left[\mlp{1/2}-\ml{1/2}\right].
    \label{eq:order_sqrt_characteristics_BL}
\end{align}
Given we are moving with a characteristic \eqref{eq:characteristic definition}, the terms in \Cref{eq:order_sqrt_characteristics_BL} involving $\madv$ cancel (validating the asymptotic scalings employed). Then, imposing the method of multiple scales constraint of periodicity on the short scale, we deduce from \Cref{eq:order_sqrt_characteristics_BL} that the solution is also independent of $\dvar$ at this order i.e.
\begin{equation}\label{eq:next_order_equality}
    \ml{1/2}(\tilde \cvar,t)=\madv^{(1/2)}(\tilde \cvar,t).
\end{equation}

At $\mathcal{O}(\eps)$ in \Cref{eq:expansion in zero advection BL PDE} we have
\begin{align}
    \frac{\partial \madv}{\partial t}= &\left(\pakdownstar1-\pakupstar1\right)\frac{\pa}{\pa \tilde \cvar}\left(\tilde \cvar\,\madv\right)+\left(\frac{\kupstar}{2}+\frac{\kdownstar}{2}\right)\frac{\pa^2\madv}{\pa \tilde \cvar^2}-\beta_c\madv\nonumber \\
    &+\kupstar\left[\mlm{1}-\ml{1}\right]+\kdownstar\left[\mlp{1}-\ml{1}\right],
\end{align}
again using the definition of $\charac$ (\Cref{eq:characteristic definition}).
Applying the Fredholm Alternative Theorem (see \Cref{sec:outer regions} and Appendix \ref{app:FAT}) leads to the following solvability condition
\begin{subequations}
\begin{align}
    \frac{\pa \madv}{\pa t}=\frac{\pa}{\pa \tilde \cvar}\left(\diffcoef\frac{\pa\madv}{\pa \tilde \cvar}-\velatc\tilde \cvar\madv\right)-\beta_c\madv,
    \label{eq:characteristics BL solvability condition}
\end{align}
where
\begin{align}
    \diffcoef(t)=\frac{1}{2}\left(\kupstar+\kdownstar\right), \quad \velatc(t)=\frac{\pa K}{\pa \cvar}(\charac(t),t)=\pakupstar1-\pakdownstar1.
\end{align}
\end{subequations}
\Cref{eq:characteristics BL solvability condition} is a parabolic advection-diffusion PDE and governs the leading-order dynamics of the structured population in the inner (moving) boundary layer around any characteristic at $\cvar=\charac(t)$. Thus, an inner continuum region can emerge in which the apparent discontinuity in the outer continuum approximation is resolved. \changed{Since $\velatc<0$, the advection in \Cref{eq:characteristics BL solvability condition} directs the population towards $\tilde\cvar=0$ from both sides, tending to focus it around this point. Effective diffusion, with coefficient $\diffcoef$, opposes this focusing and balances compression by the converging characteristics over a width $\order(\sqrteps)$ in the original structure variable. We note that here we have used the assumption that $\pa K/\pa\cvar$ is bounded away from zero. If $\velatc=0$ for some $s=\charac$, then the scalings of the inner region around the corresponding characteristic would be different, with precise scalings depending on the specific degeneracy of this point.} The appropriate boundary conditions for \Cref{eq:characteristics BL solvability condition} are obtained by matching with the outer solutions as they approach this inner region. 

%\changed{Another location of interest is the stagnation point %$\cvar=\scrit(t)$, where 
%\begin{align}\label{eq:stagnation point definition}
%    K(\scrit(t),t)=0
%\end{align}
%and towards which the characteristics converge. When the stagnation point moves, it does not generally follow a characteristic, since the local advection velocity vanishes at $\cvar=\scrit(t)$. Consequently, the characteristics may lag behind its instantaneous position, with their displacement depending on the history of its motion. At steady state, however, the characteristics converge towards the stationary stagnation point, where the steady-state outer problem becomes singular. The boundary layers associated with these characteristics therefore coalesce into a single inner region, which regularises the singular behaviour of the outer solution.}

\changed{Inner layers are required where the outer approximation develops discontinuities or other singular behaviour. As discussed in \Cref{sec:effective BCs}, such discontinuities may arise along the characteristics passing through $(\cvar,t)=(0,0)$ and $(1,0)$ when the boundary and initial conditions are incompatible. These correspond to the left and right wavefronts (regions IN1 and IN2 in \Cref{fig:initial_asymptotic_structure}a), respectively. More generally, discontinuities in the initial data would also propagate along characteristics and give rise to corresponding inner layers. Since we assume a smooth initial profile, we do not consider those additional layers here; however, their analysis would follow the same procedure.}

%\subsubsection{Inner boundary layers at the wavefronts}\label{sec:wavefront}

\begin{figure}[hbt!]
\centering
    \includegraphics[width=1\linewidth]{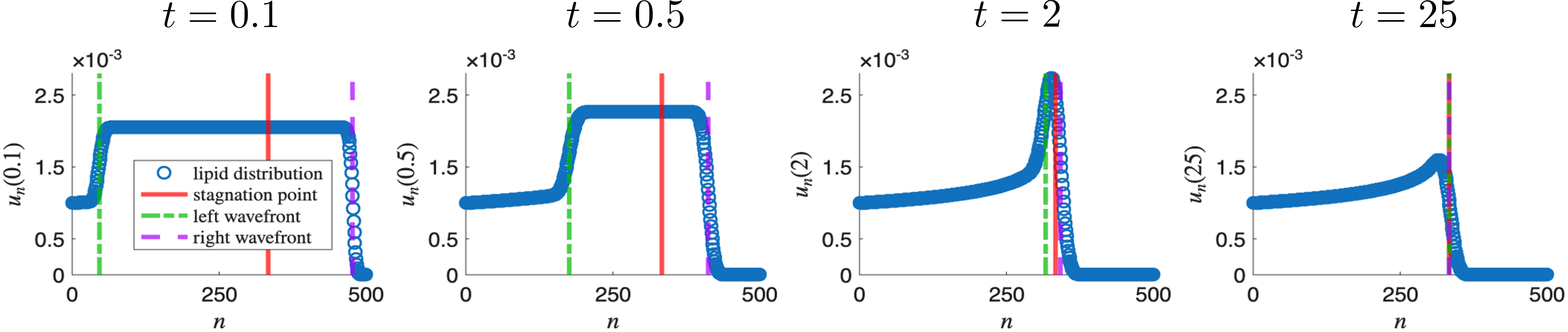}
    \caption{Distribution $\dfun_{\dvar}(t)$ for $\dvar=0,1,\dots,\lmax$ obtained by solving \Cref{eq:dimensional_structured_model} for $\lmax=500$ with linear functional forms given by \Cref{eq:linear_functions}. \changed{The red line shows the position of the stagnation point (\Cref{eq:example stag point location}), the green and purple lines respectively show how the positions of the left wavefront $\cvar=\sstar(t)$ (given by \Cref{eq:characteristic definition,eq:wavefronts ICs} with $\charac=\sstar$) and right wavefront $\cvar=\sstarr(t)$ (given by \Cref{eq:characteristic definition,eq:wavefronts ICs} with $\charac=\sstarr$) change over time.} The dynamics are shown at the time points $t=0.1,0.5,2$ and $25$. The initial conditions are taken to be $\dfun_{\dvar}(0)=\lmax^{-1}$ and we set $\constup=1,\constdown=0.5,\gamma=1.25$.}
    \label{fig:distribution_with_left_right_wavefronts}
\end{figure}

\changed{We define the positions of the left and right wavefronts in the outer system to be at $\cvar=\sstar(t)$ and $\cvar=\sstarr(t)$. Then, their positions are governed by \Cref{eq:characteristic definition} with $\charac=\sstar$ and $\charac=\sstarr$, respectively, and with initial conditions 
\begin{equation}\label{eq:wavefronts ICs}
    \sstar(0)=0, \qquad \sstarr(0)=1.
\end{equation}}
Before the wavefront layers overlap, the characteristic projections through $(0,0)$ and $(1,0)$ partition the $(\cvar,t)$-plane into three regions, separated by apparent discontinuities. The characteristics in $0<\cvar<\sstar(t)$ originate from the left boundary, those in $\sstar(t)<\cvar<\sstarr(t)$ emanate from the initial data, and those in $\sstarr(t)<\cvar<1$ originate from the right boundary. These correspond to regions O1, O2 and O3 in \Cref{fig:initial_asymptotic_structure}a, respectively.
\Cref{fig:distribution_with_left_right_wavefronts} illustrates this behaviour in numerical simulations for the linear case discussed in \Cref{sec:numerical simulations}. The wavefronts are the locations in structure space where the population distribution changes rapidly.

\changed{With $\charac=\sstar$ and $\charac=\sstarr$, \Cref{eq:characteristics BL solvability condition} governs the leading-order dynamics in the left and right wavefront layers (regions IN1 and IN2 in \Cref{fig:initial_asymptotic_structure}a), respectively, smoothing the apparent discontinuities in the outer approximation. Its far-field behaviour is determined by matching with regions O1 and O2 for the left layer, and O2 and O3 for the right layer.
We also note that, if the transition rates $\kupdown(s,t)$ approach time-independent limits, $\sstar$ and $\sstarr$ converge to the stagnation point of the limiting advection velocity as $t\to\infty$. The two versions of \Cref{eq:characteristics BL solvability condition}, with $\charac=\sstar$ and $\charac=\sstarr$, therefore coincide in this limit. The corresponding inner boundary layers coalesce into a single steady-state asymptotic region (region IN4 in \Cref{fig:initial_asymptotic_structure}b) centred at the stagnation point.}

\section{Example: early-stage atherosclerosis} \label{sec:example}

We now demonstrate the implications of our analysis for a specific application: the dynamics of macrophage-lipid interactions during the early stages of atherosclerosis \cite{chambers2023new}. During early atherosclerosis, low-density lipoprotein (LDL) particles penetrate the artery wall, eliciting an inflammatory response and promoting the recruitment of macrophages that engulf LDL \cite{boren2020low,tabas2016macrophage}. In parallel, macrophages offload excess cholesterol to high-density lipoprotein (HDL), which transports it out of the lesion \cite{back2019inflammation}. 

\subsection{Continuum approximation}
The lipid-structured model in \cite{chambers2023new} decomposes the macrophages into discrete compartments characterised by lipid load. Macrophages move between compartments when they take up and offload lipid. We denote by $\dfun_{\dvar}(t)\geq0$ the dimensionless concentration of macrophages in the atherosclerotic plaque with $\dvar=0,1,\dots,\lmax$ units of lipid at time $t$, where $\lmax>0$ defines the macrophage lipid capacity. The functions $\klup{\dvar}(t),\kldown{\dvar}(t)$ represent the rates of lipid uptake and offloading, respectively, and their functional forms, along with those for the recruitment and removal functions $f(t),g(t),\betal{\dvar}(t)$, are given in \Cref{eq:linear_functions}. Writing $\lmax=\eps^{-1}$, \Cref{eq:ks_independence_of_discrete_variable} gives
\begin{align}
    \kup(\eps\dvar,t)=\constup(1-\eps\dvar), && \kdown(\eps\dvar,t)=\constdown\eps\dvar, && \beta(\eps\dvar,t)=\gamma.\label{eq:example functional forms no discrete variable}
\end{align}
The dynamics in the outer regions are governed by \Cref{eq:secularity_condition_outer} which, for this application, becomes
\begin{equation}
    \frac{\pa \dfun}{\pa t}=\frac{\pa}{\pa \cvar}\left[\left(\constdown \cvar-\constup(1-\cvar)\right)\dfun \right] - \gamma \dfun,
    \label{eq:outer PDE example}
\end{equation}
with the following boundary conditions, obtained from \Cref{eq:left_BC,eq:right_BC}:
\begin{align}
    \dfun(0^+,t)=\frac{1}{2\constup}, && \dfun(1^-,t)=0.\label{eq:BCs for example}
\end{align}
In the inner region around the left wavefront $\cvar=\sstar(t)$, \Cref{eq:characteristics BL solvability condition} yields the macrophage dynamics via the following advection-diffusion PDE
\begin{align}
    \frac{\pa \mfront}{\pa t}=\frac{1}{2}\left[\constup(1-\sstar)+\constdown\sstar\right]\frac{\pa^2\mfront}{\pa \tilde \cvar^2}+(\constup+\constdown)\frac{\pa}{\pa \tilde \cvar}\left(\tilde \cvar\mfront\right)-\gamma \mfront,
\end{align}
with an equivalent PDE around $\cvar=\sstarr(t)$.
\changed{The stagnation point is located where $K(s,t)=0$, which, given \Cref{eq:example functional forms no discrete variable}, is at}
\begin{equation}\label{eq:example stag point location}
    \scrit=\frac{\constup}{\constup+\constdown}.
\end{equation}
\changed{\Cref{fig:distribution_with_left_right_wavefronts}, which presents the lipid distribution, along with the locations of the inner boundary layers $\cvar=\sstar(t),\sstarr(t)$ and the stagnation point $\cvar=\scrit$, shows that for $t=0.1,0.5$ the two regions of rapid variation are well separated, but they both move inwards towards the stagnation point. For $t = 2, 25$, the wavefronts merge into one region of rapid variation (IN4), which receives boundary data from outer regions O1 and O3.}

\subsection{Numerical validation}

\begin{figure}[ht!]
\centering
\includegraphics[width=0.5\linewidth]{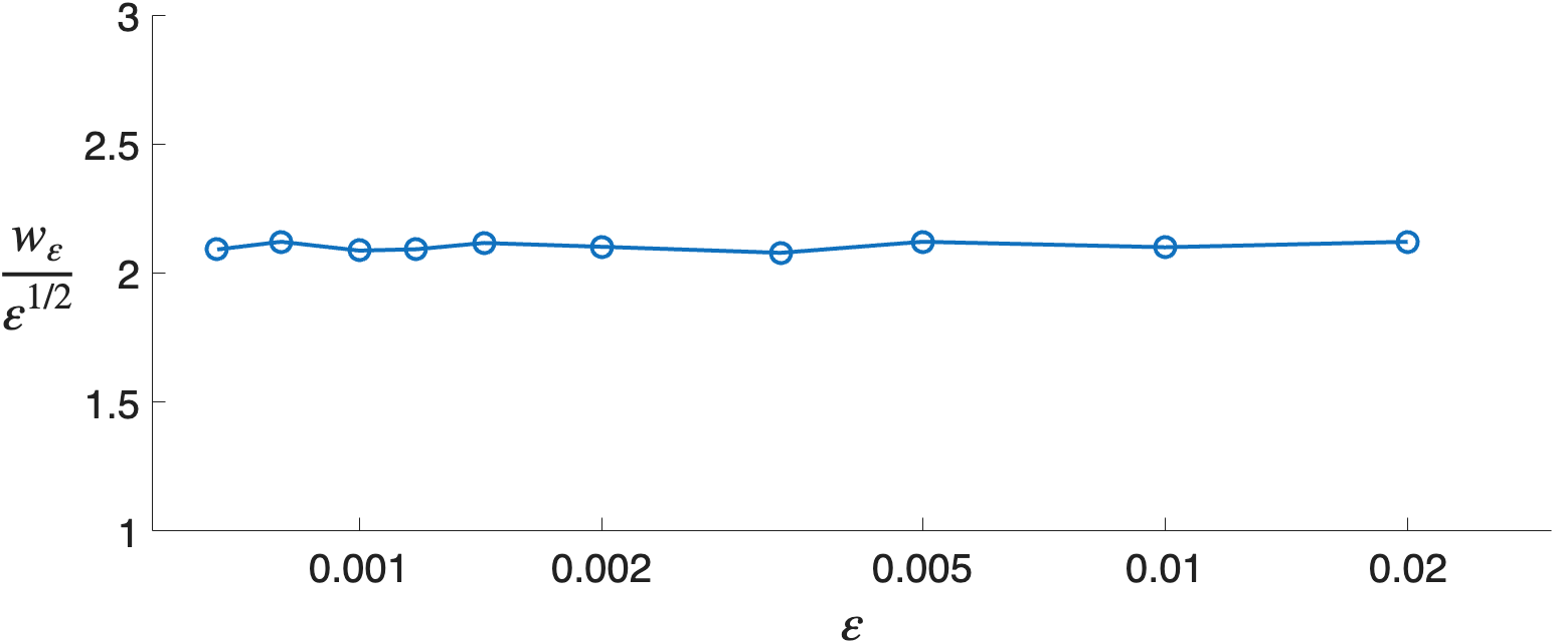}
\caption{Plot of $w_{\eps}/\eps^{1/2}$, where $w_{\eps}$ is the width of the inner boundary layer at approximate steady state ($t=25$), defined by \Cref{eq:BL width}. The width is numerically computed by identifying the beginning and end of the boundary layer as defined in \Cref{eq:start of BL analytical definition,eq:end of BL analytical definition}, respectively. We set $\constup=1,\constdown=0.5,\gamma=1.25$.}
\label{fig:BL_width_vs_eps}
\end{figure}

\changed{In \Cref{fig:BL_width_vs_eps} we numerically validate the scaling of the inner boundary layer around a characteristic, as identified in \Cref{eq:inner_variables_definition_characteristic}. Since the advection velocity is time-independent in this example, all characteristics converge to the stagnation point $\cvar=\scrit$ (defined in \Cref{eq:example stag point location}) as $t\to\infty$, giving rise to a single inner boundary layer at steady state (region IN4 in \Cref{fig:initial_asymptotic_structure}). Numerically defining the extent of the boundary layer requires some care, since the solution to the outer problem defined by \Cref{eq:outer PDE example,eq:BCs for example} can diverge as $\cvar \to \scrit^{-}$ when $\constup+\constdown>\gamma$. For finite $\lmax$ this divergent outer behaviour is regularised by the change in scaling within the boundary layer. The boundary layer can therefore be defined from the left as the region near the maximum of $\dfun_{\dvar}$, and from the right as the region where $\dfun_{\dvar}$ departs from the right-hand outer solution of zero.}

We therefore estimate the width of the inner boundary layer by first identifying the value of $\dvar$ at which $\dfun_{\dvar}(25)$ attains its maximum value, writing
\begin{equation}
    \dvar_*:=\operatorname*{arg\,max}_{\dvar\in\{0,1,\dots,\lmax\}}\dfun_{\dvar}(25),
    \qquad
    \dfun_{\dvar_*}^{\max}:=\dfun_{\dvar_*}(25).
\end{equation}
The start of the boundary layer is defined as the rightmost point $\dvar=\dvar_{\text{start}}<\dvar_*$ at which the distribution is at most 98\% of this maximum, such that
\begin{align}
    \dvar_{\rm start}=\max\{\dvar<\dvar_*:\dfun_\dvar(25)\le0.98\dfun_{\dvar_*}^{\max}\},
    \qquad \dfun_{\dvar_{\text{start}}}(25)\leq0.98\dfun_{\dvar_*}^{\max}.\label{eq:start of BL analytical definition}
\end{align}
The end of the boundary layer is defined as the leftmost point $\dvar=\dvar_{\text{end}}>\dvar_*$ at which the distribution has decayed to 2\% of its maximum, i.e. 
\begin{equation}
    \dvar_{\rm end}=\min\{\dvar>\dvar_*:\dfun_\dvar(25)\le0.02\dfun_{\dvar_*}^{\max}\},
    \qquad
    \dfun_{\dvar_{\text{end}}}(25)\leq0.02\dfun_{\dvar_*}^{\max}.\label{eq:end of BL analytical definition}
\end{equation}
Then, the width of the boundary layer is defined as
\begin{equation}
    w_{\eps}:=\eps(\dvar_{\text{end}}-\dvar_{\text{start}}).\label{eq:BL width}
\end{equation}
\Cref{fig:BL_width_vs_eps} shows a plot of $w_{\eps}/\eps^{1/2}$ as $\eps$ varies. As expected from the inner scaling given by \Cref{eq:inner_variables_definition_characteristic}, the plot is approximately constant in $\eps$ i.e. the boundary layer width scales linearly with $\eps^{1/2}$. 

\begin{figure}[ht!]
\centering
\includegraphics[width=1\linewidth]{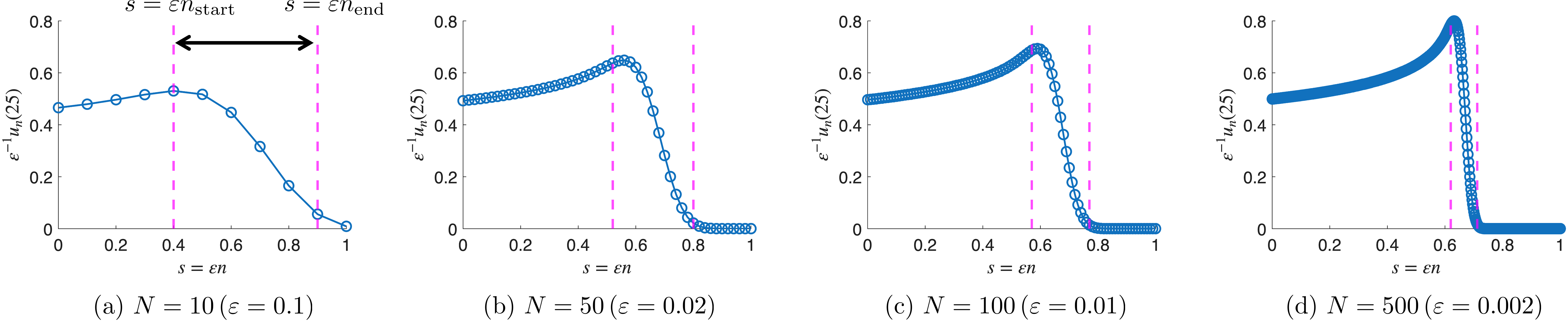}
\caption{Macrophage distribution with linear functional forms at approximate steady state ($t=25$), $\dfun_{\dvar}(25)$. We plot the rescaled distributions $\eps^{-1}\dfun_{\dvar}$ over $\cvar=\eps\dvar$. The pink vertical dashed lines mark the beginning ($\cvar=\eps\dvar_{\text{start}}$) and end ($\cvar=\eps\dvar_{\text{end}}$) of the inner boundary layer, as given by \Cref{eq:end of BL analytical definition,eq:start of BL analytical definition}, for different values of $\lmax$ ($\eps$). We set $\constup=1,\constdown=0.5,\gamma=1.25$.}
\label{fig:distribution_with_BL_position}
\end{figure}

In \Cref{fig:distribution_with_BL_position}, we plot the macrophage distribution, indicating the position of the beginning and end of the inner boundary layer from this procedure with dashed vertical lines. As $\lmax$ increases, the boundary layer shrinks, as expected. 

\subsection{Comparison of discrete and continuum models}

We compare the composite continuum approximation and the solution of the original discrete model in various asymptotic regions, by considering a case for which the system is at steady state. For simplicity, we focus on a specific choice of parameters for which the steady-state solutions in all outer regions are constant. We choose
\begin{equation}
    \gamma=\constup+\constdown.\label{eq:parameters for discrete vs continuum}
\end{equation}

In the outer regions we solve \Cref{eq:outer PDE example} with $\frac{\pa\mouter}{\pa t}=0$. As $t\to\infty$, we have $\sstar(t),\sstarr(t)\to\scrit$. The steady-state solution $\msteadyouter(\cvar)=\lim_{t\to\infty}\mouter(\cvar,t)$ is given by
\begin{equation}
    \msteadyouter(\cvar)=
    \begin{cases}
    \frac{1}{2\constup}\left(1-\frac{\constup+\constdown}{\constup}\cvar\right)^{-1+\frac{\gamma}{\constup+\constdown}}, &\cvar\in[0,\scrit), \\
    %\frac{1}{\lmax}\mathrm{e}^{(\constup+\constdown-\gamma)t}, &\cvar\in(\sstar,\scrit)\,\,\text{and}\,\,\cvar\in(\scrit,\sstarr),\\
    0, &\cvar\in(\scrit,1],
    \end{cases}
    \label{eq:analytical sol outer}
\end{equation}
which is constant in $\cvar$ for our parameter choice given by \Cref{eq:parameters for discrete vs continuum}. Here, $\cvar\in[0,\scrit)$ corresponds to region O1, and $\cvar\in(\scrit,1]$ to region O3 in \Cref{fig:initial_asymptotic_structure}.

At steady state, the dynamics in the inner boundary layer are given by
\begin{equation}
    \frac{1}{2}\left[\constup(1-\scrit)+\constdown\scrit\right]\frac{\dd^2\msteadyinner}{\dd \tilde \cvar^2}+(\constup+\constdown)\tilde\cvar\frac{\dd\msteadyinner}{\dd \tilde \cvar}+(\constup+\constdown-\gamma) \msteadyinner=0,
    \label{eq:inner ODE steady state specific case}
\end{equation}
\changed{where $\msteadyinner(\tilde \cvar)=\lim_{t\to\infty}\madv(\tilde \cvar,t)$ is the steady-state distribution inside the inner boundary layer and $\tilde \cvar=\sqrt{\eps}\left(\dvar-\frac{\scrit}{\eps}\right)\in(-\infty,\infty)$, as given by \Cref{eq:inner_variables_definition_characteristic} with $\charac=\scrit$.} With the choice of parameters given by \Cref{eq:parameters for discrete vs continuum}, we deduce from \Cref{eq:analytical sol outer} that the matching conditions for $\msteadyinner$ are
\begin{align}
    \msteadyinner\to\frac{1}{2\constup}\,\,\text{as}\,\,\tilde \cvar\to-\infty, && \msteadyinner\to0\,\,\text{as}\,\,\tilde \cvar\to\infty.
    \label{eq:inner BCs specific case}
\end{align}
Solving \Cref{eq:inner ODE steady state specific case,eq:inner BCs specific case} with $\gamma=\constup+\constdown$ yields
\begin{equation}
    \msteadyinner(\tilde \cvar)=\frac{1}{4\constup}\mathrm{erfc}\left(\tilde \cvar\sqrt{\frac{\constup+\constdown}{\constup(1-\scrit)+\constdown\scrit}}\right),\label{eq:analytical sol inner tilde \cvar}
\end{equation}
where $\mathrm{erfc}(z)=1-\mathrm{erf}(z)$ is the complementary error function. Rewriting \Cref{eq:analytical sol inner tilde \cvar} in terms of the outer variable $\cvar$, using $\cvar=\eps\dvar$ and \Cref{eq:inner_variables_definition_characteristic}, gives
\begin{equation}
    \msteadyinner(\cvar)=\frac{1}{4\constup}\mathrm{erfc}\left[(\cvar-\scrit)\left(1+\frac{\constdown}{\constup}\right)\sqrt{\frac{\lmax}{2}\frac{\constup}{\constdown}}\right],
    \label{eq:analytical solution inner}
\end{equation}
and $\scrit$ is given by \changed{\Cref{eq:example stag point location}.}
Finally, the composite solution in $\cvar\in[0,1]$ is given by $\dfun^*(\cvar)\equiv\msteadyinner(\cvar)$.

\begin{figure}[ht!]
\centering
\includegraphics[width=1\linewidth]{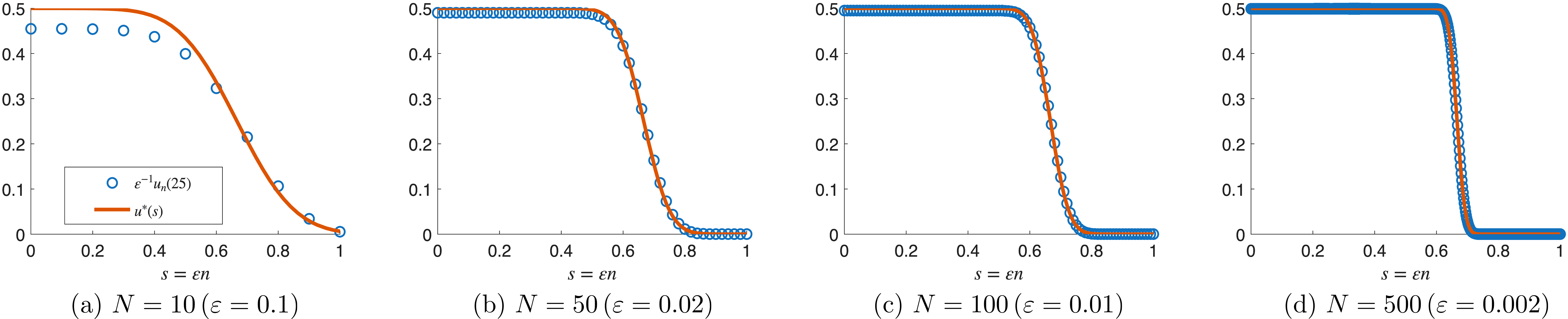}
\caption{The rescaled solution of the discrete model $\eps^{-1}\dfun_{\dvar}(25)$ at $t=25$ (blue dots) is compared to the steady-state solution of the continuum model $\dfun^*(\cvar)$ (solid red curve), for different values of $\lmax$. Parameter values: $\constup=1,\constdown=0.5,\gamma=\constup+\constdown$.}
\label{fig:discrete vs continuum}
\end{figure}

In \Cref{fig:discrete vs continuum} we compare the composite continuum solution $\dfun^*(\cvar)$ to the solution of the discrete model \eqref{eq:dimensional_structured_model} at $t=25$ (approximate steady state), $\dfun_{\dvar}(t=25)$, for different values of $\lmax$. We find that the discrete and continuum models are in excellent agreement, with
increased accuracy as $\lmax$ increases, as expected. 

\section{Discussion and conclusion}\label{sec:conclusion}

We present a systematic discrete multiscale framework for the derivation and analysis of continuum approximations to discretely structured mathematical models. Continuum limits of discrete systems are commonly studied in the literature (see e.g. \cite{chambers2023new,lai2025impact,chaplain2020bridging,stace2020discrete,kory2024discrete,wattis2020mathematical,pillay2017modeling,barry2022continuum,kory2024discrete2}); our analysis addresses unresolved ambiguities about truncation order, boundary conditions, degeneracies, and discontinuities that often arise in approaches based solely on classic Taylor series expansions. The continuum description we derive is mathematically consistent and uniformly valid across the continuum structure space, and our analysis also identifies where a truncated continuum description fundamentally cannot approximate the discrete model. By constructing solutions in different asymptotic regions and matching them consistently, we ensure that the discrete-to-continuum approximation captures the essential dynamics in the appropriate regions of structure space. \changed{Our framework does not address whether a discrete or a continuum description is more appropriate for a given biological system; this is a modelling choice, typically led by the biology. Rather, given a discrete structured model, it shows when and where consistent continuum approximations are appropriate, and how to derive them.
}

\changed{We studied a population model structured into discrete subpopulations.} To analyse this, we introduced a discrete short scale alongside a continuum long scale using the method of multiple scales applied to discrete systems. This scale separation captures microscopic heterogeneity in the population distribution while also allowing for macroscopic variations. Under the assumption that the transition rates between neighbouring subpopulations vary over the long scale, we showed that local transitions dominate the leading-order behaviour in the outer regions, where the dynamics can be described by a hyperbolic PDE. The macroscale advection observed in these regions is consistent with the continuum limits derived by Chambers et al. \cite{chambers2023new} and Lai et al. \cite{lai2025impact} using Taylor expansion-based approaches. However, this Taylor expansion approach breaks down near endpoints and inner boundary layers. Our multiscale framework provides a formal identification of where in structure space these continuum representations are valid, where they break down and, in the latter scenario, a systematic derivation of the correct reduced representations to impose. This reinforces the idea that population-level transport phenomena can be derived consistently from microscopic transition mechanisms, rather than introduced phenomenologically.

To describe the dynamics near points of apparent discontinuities in the outer regions, \changed{we employed a boundary layer analysis by first rescaling the discrete system within the corresponding inner region and subsequently upscaling it.} \changed{Regions of rapid variation arise from discontinuities between neighbouring characteristics, with the resulting inner layers governed by advection-diffusion equations on the $\order(1/\sqrt{\lmax})$ scale, where $\lmax\gg1$ is the number of structure classes.} Across these rapid transition regions, diffusive effects, proportional to the rate of structure-changing interactions \(\klup{\dvar}+\kldown{\dvar}\), regularise the apparent discontinuities in the outer solutions, ensuring smooth transitions in the population structure. 

\changed{We also clarify the truncation-order ambiguity that can complicate Taylor-based continuum approximations. In our approach, the order of each continuum equation is set by the dominant balance in its region rather than chosen \emph{a priori}. The self-consistent leading-order system we derive consists of: first-order hyperbolic PDEs in the outer regions; second-order parabolic PDEs in the inner regions, which resolve the apparent discontinuities in the outer solution; and fundamentally discrete equations in the boundary layers at each end of the structure domain, where no truncated continuum approximation holds.} Although intersection of boundary layers could produce more complex inner structures (e.g. regions B3, B4, IN3, IN4 in \Cref{fig:initial_asymptotic_structure}b), our focus here was on the principal balances that determine dominant macroscale behaviour. 

A key outcome of this analysis is that the boundary layers at the edges of the structure domain retain an inherently discrete character. That is, while the continuum limit accurately captures bulk behaviour, the boundary dynamics remain governed by transitions at the scale of the discrete variable. By explicitly analysing these regions, we derived consistent matching conditions between discrete inner and continuum outer solutions, yielding consistent PDEs with systematically derived boundary conditions. This approach allows one to resolve a technical difficulty in continuum limit derivations -- namely, how to systematically determine boundary conditions that remain valid across scales, formalising the mathematical foundations laid by earlier heuristic treatments (e.g. \cite{chambers2023new,lai2025impact}).

% In \Cref{sec:example}, we applied the framework to a simplified model of early atherosclerosis, based on \cite{chambers2023new}, yielding excellent agreement between our upscaled continuum model and the original discrete simulations. This application illustrates how the method can be used to derive consistent continuum models from biologically relevant structured systems.

For simplicity, we assumed throughout that recruitment occurs only at domain boundaries. It is possible to generalise this assumption to allow for distributed recruitment across subpopulations. Under slow variation of recruitment rates, this would introduce an additional source term in the continuum equations. Such an extension would enable the modelling of biological processes such as \changed{immigration from an external supply or differentiation from another population \cite{calsina2016structured}.} Indeed, this study neglected proliferation within the structured population. Although biologically significant, proliferation introduces technical challenges in multiscale analysis. Extending the discrete multiple-scales framework to incorporate proliferation would enable the systematic upscaling of non-local growth terms. 

The analysis relies critically on the assumption of slow, monotonic variation in the transition rates. Relaxing this macroscale monotonicity would introduce additional asymptotic complexity, as the forward and backward transition rates could intersect multiple times within the structure domain, producing several stagnation points and hence multiple inner boundary layers. This could lead to technically challenging inner-region analyses, with information potentially propagating from repelling stagnation points into the outer regions (in contrast to the flow of information in our model). An interesting extension to the present work would be to consider transition rates with a periodic short-scale heterogeneity, combining the current analysis with the discrete multiple-scales homogenisation framework of \cite{chapman2017effective}, where continuum approximations of random walks on periodic lattices are derived. Such an extension would lead to interesting mathematical challenges within the corresponding inner regions.

Relaxing the constraint of monotonic variation in transition rates can also lead to both transition rates becoming asymptotically small near an endpoint (such as in \cite{lai2025impact}). In such cases, the analysis we conducted at the boundaries here is no longer valid, and interesting mathematical subtleties may arise. This highlights how discrete boundary effects could qualitatively alter the continuum behaviour, suggesting a rich mathematical structure that warrants further exploration.

Finally, a natural extension to this work involves including spatial effects, which may depend on the structural class (e.g. structure-dependent diffusion \cite{ridgway2023motility}). Accounting for these effects would provide a more comprehensive multiscale representation of coupled spatial and structural dynamics.

To conclude, our results demonstrate how the method of multiple scales applied to discrete systems can construct uniformly valid continuum approximations of structured population models. The framework clarifies ambiguities inherent in Taylor expansion methods, systematically derives continuum boundary conditions from discrete dynamics, and captures the smooth resolution of apparent discontinuities. More broadly, it demonstrates how systematic multiscale analysis can connect discrete microscopic biological organisation with continuum-level macroscopic dynamics.

\appendix

\section{Fredholm alternative theorem for linear difference operators}\label{app:FAT}

\changed{We consider the Fredholm alternative theorem for linear difference operators acting on $p$-periodic sequences. In the context of the multiple-scales analysis used in this paper, this periodicity arises from the periodicity constraint obtained on the short discrete scale at leading order (see Step 5 in \Cref{sec:outer regions}).}

The $\order(\eps)$ \Cref{eq:outer_region_order1_eq} can be written in terms of a linear difference operator,
\begin{subequations}
\begin{align}
(L\mouter^{(1)})_\dvar = h[\mouter],
\end{align}
where, in the outer regions (where $\dvar,\eps^{-1}-\dvar\gg1$), $L$ is defined as
\begin{align}
(Lv)_\dvar := \kdown v_{\dvar+1} - (\kup + \kdown) v_\dvar + \kup v_{\dvar-1}, \quad
h[\mouter] := \frac{\partial \mouter}{\partial t}
+\frac{\partial}{\partial \cvar}\left(K\mouter\right)
+\beta\mouter.
\end{align}
\end{subequations}
To formally apply the Fredholm alternative theorem, we introduce the shift operator $E$, defined by
\begin{equation}
    (Ev)_\dvar=v_{\dvar+1}.
\end{equation}
Then we have
\begin{equation}
    L=\kdown E-(\kup+\kdown)I+\kup E^{-1},\label{eq:L operator}
\end{equation}
where $I$ is the identity operator. For $p$-periodic sequences, a suitable inner product is
\begin{align}
    \langle v,w\rangle=\sum_{\dvar=0}^{p-1}v_\dvar\overline{w_\dvar}.
\end{align}
In order to obtain a solvability condition via the Fredholm alternative, we must first determine the adjoint operator via
\begin{equation}
    \langle Ev,w\rangle
    =\sum_{\dvar=0}^{p-1} v_{\dvar+1}\overline{w_\dvar}
    =\sum_{\dvar=1}^{p} v_\dvar \overline{w_{\dvar-1}}
    =\sum_{\dvar=0}^{p-1} v_\dvar\overline{w_{\dvar-1}}
    =\langle v,E^{-1}w\rangle,
\end{equation}
where the relabelling of indices follows from $p$-periodicity, so that $v_p=v_0$ and $w_{p-1}=w_{-1}$. Therefore, $E^*=E^{-1}$.

In particular, for $L$ given by \Cref{eq:L operator}, we have
\begin{equation}
    L^*=\kdown E^{-1}-(\kup+\kdown)I+\kup E.\label{eq:adjoint operator}
\end{equation}
The Fredholm alternative states that a nontrivial solution $y$ to $L^* y = 0$ yields the solvability condition
\begin{equation}
    \langle y,h[\mouter]\rangle = 0.
\end{equation}
From \Cref{eq:adjoint operator}, we find that the adjoint operator $L^*$ has kernel
\begin{equation}
    \ker L^*=\text{span}\{1,(\kdown/\kup)^\dvar\},
\end{equation}
for $\kup\neq\kdown$, of which only the constant solution is $p$-periodic in general. In our analysis, this is therefore the solution compatible with the method of multiple scales constraint of strict periodicity on the short scale. Hence, the Fredholm alternative provides the solvability condition
\begin{equation}
    0 = \langle y,h[\mouter]\rangle
    = \langle1,h[\mouter]\rangle
    = p\, h[\mouter].
\end{equation}

\bibliography{journal_abbreviations,references}
\bibliographystyle{abbrvnat}
\end{document}

%% file: macros.tex
\newcommand{\dvar}{n}
\newcommand{\cvar}{s}
\newcommand{\pa}{\partial}
\newcommand{\eps}{\varepsilon}
\newcommand{\dd}{\mathrm{d}}
\newcommand{\sstar}{\ell}

\newcommand{\lmax}{N}

\newcommand{\klup}[1]{k_{#1}^+}
\newcommand{\kup}{k^+}
\newcommand{\kldown}[1]{k_{#1}^-}
\newcommand{\kdown}{k^-}
\newcommand{\minitl}[1]{U_{#1}}
\newcommand{\minit}{U}
\newcommand{\klupdown}[1]{k_{#1}^\pm}
\newcommand{\order}{\mathcal{O}}
\newcommand{\kupstar}{k^+_c}
\newcommand{\kdownstar}{k^-_c}
\newcommand{\kupdownstar}{k^{\pm}_c}
\newcommand{\pakupstar}[1]{\partial^{#1}k^+_c}
\newcommand{\pakdownstar}[1]{\partial^{#1}k^-_c}
\newcommand{\pakupdownstar}[1]{\partial^{#1}k^{\pm}_c}
\newcommand{\sqrteps}{\sqrt{{\varepsilon}}}
\newcommand{\kupdown}{k^{\pm}}
\newcommand{\mfront}{\dfun_{\sstar}}
\newcommand{\betal}[1]{\beta_{#1}}
\newcommand{\sstarr}{r}

\newcommand{\scrit}{s_c}
\newcommand{\madv}{\dfun_c}

\newcommand{\constup}{k_f}
\newcommand{\constdown}{k_b}
\newcommand{\msteadyouter}{u^*_{\text{outer}}}
\newcommand{\msteadyinner}{u^*_{\text{inner}}}

\newcommand{\dfun}{u}
\newcommand{\ml}[1]{\dfun_{\dvar}^{(#1)}}
\newcommand{\mlm}[1]{\dfun_{\dvar-1}^{(#1)}}
\newcommand{\mlp}[1]{\dfun_{\dvar+1}^{(#1)}}
\newcommand{\mouter}{\dfun}
\newcommand{\diffcoef}{D_c}
\newcommand{\charac}{c}
\newcommand{\velatc}{K'_{\charac}}